\documentclass[12pt]{iopart}
\usepackage{dsfont}
\usepackage[singlelinecheck=off]{caption}
\usepackage{graphicx}
\usepackage{tabularx}
\usepackage[utf8x]{inputenc} 

\usepackage{epsfig}
\usepackage{ucs}
\usepackage{color}
\usepackage{units}
\usepackage{threeparttable}
\usepackage{latexsym}
\usepackage{amsfonts}
\usepackage{amssymb}
\usepackage{textcomp}

\newcommand{\bef}{\begin{figure}}
\newcommand{\eef}{\end{figure}}
\newcommand{\ec}{\end{center}}

\newcommand{\be}{\begin{eqnarray}}
\newcommand{\ben}{\begin{enumerate}}

\newcommand{\ra}{\rangle}
\newcommand{\een}{\end{enumerate}}
\newcommand{\ee}{\end{eqnarray}}

\begin{document}
\title[Divergence of time-dependent perturbation theory]{
On the divergence of time-dependent perturbation theory applied to laser-induced molecular transitions: Analytical calculations for the simple algorithm}

\author{
Klaus Renziehausen }

\address{
Universit\"at W\"urzburg \\
Institut f\"ur Physikalische und Theoretische Chemie\\
Emil-Fischer-Straße 42, 97074 W\"urzburg, Germany 
\vspace{0.5cm}\\
Email: ksrenzie@phys-chem.uni-wuerzburg.de}
\date{\today}

\begin{abstract}
\noindent
Shaped laser pulses are a powerful tool to induce population transfer between electronic molecular states, and
time-dependent perturbation theory is suitable for a description of such a transfer in weak external fields. The application of perturbation theory in numerical simulations of field matter interactions can lead to divergences. In a recent paper [K. Renziehausen et. al.  {\em J. Phys. B: At. Mol. Opt. Phys.}, 42:195402, 2009] we explained that the arising error in the norm of the wave function can be split into two parts. The first part is related to numerical errors caused by the discretisation of time that is required in the simulation and can be suppressed for a sufficiently small time step or abolished for an adequate numerical implementation of perturbation theory. 
The second part may cause divergences and is associated with the perturbative expansion order. We presented numerical evidence without any analytical proof. Here we are focussing on the derivation of analytical expressions to interpret the behavior of what we have called in the above mentioned paper 'simple algorithm'. The derivation of analytical expressions for the interpretation of what we have called in the above mentioned paper 'improved algorithm' are given in another paper [K. Renziehausen. {\em arXiv}, in preparation, 2012].  
Moreover, we introduce here a gedankenexperiment to illustrate the influence of the different orders on the field-molecule interaction.
\end{abstract}
\pacs{32.80.Wr,31.15.xp,42.60.Jf,33.80.Be,31.50.-x \newline 2010 MSC: 81Q15}
\maketitle

\newpage
\section{\label{intro} Introduction} \noindent
In a recently published work \cite{Renziehausen09JPA}, we discussed in how far numerical algorithms for the implementation of perturbation theory are applicable to ultra-short laser pulse molecule interactions. This is an important issue because time-dependent perturbation theory is most commonly employed for the investigation of the interaction of atoms \cite{Faisal87} and molecules \cite{Mukamel95} with electromagnetic fields. However, the disadvantage of the application of time-dependent perturbation theory is that 
it is generally not norm-conserving. Although non-perturbative and norm-conserving algorithms can be used to solve the time-dependent Schrödinger equation, the advantage of perturbative methods is: They  
allow for a clear decomposition of multi-photon processes into contributions which stem from different orders. As an interesting example we mention time-resolved four-wave mixing spectroscopy, where one determines the third-order polarization associated with the signal emitted into a given direction \cite{Stock97ACP}. It is possible to analyse four-wave mixing experiments with non-perturbative methods \cite{Seidner95JCP,Meyer00APB,Gelin09ACR}. Nevertheless, within perturbation theory it is possible to  differentiate contributions form higher order interactions \cite{Materny00APB,Faeder01JCP,Siebert06JORA}.

In the analysis presented in \cite{Renziehausen09JPA} we used a model of two molecular electronic states (ground state ($|0\rangle$)
and exited state ($|1\rangle$)). In this model initially only one of these two states is populated. Such
conditions are realized, e.g., in pump/shaped-dump experiments which have recently been reported
\cite{Vogt06CPL,Marquetand07EPL,Dietzek07JACS}. In such processes where high intensity laser pulses interact
with molecules, it is important to understand how the results of perturbation theory converge to the exact results. 
Although this convergence behaviour, depending in detail on the chosen numerical parameters, on the molecular system and on the electric fields, cannot easily be discribed quantitatively, 
it is still possible to analyse qualitatively the general trends how the 
deviations of the norm from unity depend on the chosen parameters and the physical situation.
In our former analysis we used a simple and an improved algorithm, and we stated that the norm deviations can be decomposed into two parts, which we named the \textit{stationary orders} and the \textit{oscillatory orders}. These different parts differ in their behaviour when the following parameters are changed:   
The time step for the discretisation of time in the numerical algorithm, the order of perturbation in which the wave functions are calculated, the shape of the potentials of the two electronic states $|0\rangle$ and $|1\rangle$ and the electric field. In short the stationary orders occur only for the simple algorithm; they depend on the electric field and the time step; they converge to zero for the limit that the time step goes to zero, but they do not depend on the shape of the potentials. Thus, we concluded that norm deviations caused by the stationary orders are purely numerical and they are not related to norm deviations that are caused by the discarding of higher order interactions. In contrast to the norm deviations caused by the stationary orders the norm deviations caused by the oscillatory orders occur for both algorithms, they do not depend in leading order on the time step and thus they do not vanish in the limit of a vanishing time step. Moreover, they depend strongly on the chosen perturbation order for the wave function, the shape of the potentials of the two electronic states $|0\rangle$ and $|1\rangle$, and on the electric field. This clearly indicates that the errors are related to the perturbative expansion of the wave function.
The numerical results presented in  \cite{Renziehausen09JPA} were interpreted with the help of analytical expressions which, however, were given without proofs (in particular the equations (13) and (14) of \cite{Renziehausen09JPA} are not proven there). It is the purpose of the present paper to fill this gap for those analytical expressions, which are related to the simple algorithm. The calculations for the improved algorithm will be presented in another paper \cite{Renziehausen11JMP}. Besides the mathematical analysis of the norm deviations of the perturbative wave functions new numerical results are presented, and an interpretation in terms of a scattering gedankenexperiment are given.

The paper is organized as follows: We describe in Sec. \ref{theo} the structure of the discussed
Hamiltonian, summarize the basis of perturbation theory and show how the simple algorithm introduced in \cite{Renziehausen09JPA}  using perturbation theory for the calculation of the wave function can be derived. 
Section \ref{error_wave} contains an analytical analysis of the wave function calculated with the simple algorithm, and  
in Sec. \ref{error_norm} an analytical analysis of the norm of this wave function is given (in this section we derive the above mentioned equations (13) and (14) of \cite{Renziehausen09JPA} for the simple algorithm). Then, in Sec. \ref{Interpretation} we summarize the interpretation of the analytical results of Sec. \ref{error_norm} and provide the above mentioned gedankenexperiment. The paper is finished by a summary in Sec. \ref{sum}. 
\section{Theory \label{theo}}
\subsection{Hamiltonian} \noindent
As mentioned in the introduction, we investigate the interaction of an ultrashort laser pulse with a molecule in a model where we consider two electronic states $\vert 1\rangle$ and $\vert 0 \rangle$. The nuclear degrees of freedom are represented by a single coordinate $R$. The total Hamiltonian $\hat H(R,t)$ consists of the system Hamiltonian $\hat H_0(R)$, and the field-matter interaction term $\hat W(t)$
\begin{eqnarray}
\hspace{-1 cm} \hat H(R,t) &=& \hat H_0(R) + \hat W(t)  \nonumber \\ 
\hspace{-1 cm} &=& \left( \begin{array}{ccc} \hat T + V_1(R) & 0 \\ 0 &  \hat T + V_0(R) \end{array} \right) \label{three} 
 +  \left( \begin{array}{ccc} 0 & -\mu E(t) \\ -\mu E(t) &  0 \end{array} \right), \label{four}
\end{eqnarray}
where $\hat T$ is the kinetic energy operator, and where $V_{j}(R),~ j \in \lbrace 0,1 \rbrace$ are the potentials in the electronic states. The perturbation consists of the dipole interaction with the electric field $E(t)$ of the laser pulse and the projection $\mu$ of the transition dipole moment on the laser polarization vector. We take into account the Condon approximation and neglect the dependence of the transition dipole-moment on the nuclear coordinates. Moreover, dipole-coupling within a single electronic state is not regarded. 
As we analyse a system with two electronic states, we have to work with a two-component nuclear wave function $\vec \Psi(R,t)$ and the time-dependent Schrödinger equation reads (in atomic units):
\be
i \frac{\partial}{\partial t} 
\;\left( \begin{array}{c} \Psi_{1}(R,t) \\  \Psi_{0}(R,t) \end{array} \right)  
= \hat H(R,t) 
\;\left( \begin{array}{c} \Psi_{1}(R,t) \\  \Psi_{0}(R,t) \end{array} \right) .
\label{tdse}
\ee
Due to the structure of the perturbation operator $\hat W(t)$ for even powers of $\hat W(t)$ with 
$\eta \in 
\mathbb{ N}
$ the following equation holds:
\begin{eqnarray}
W(t)^{2\eta} &=& \mu^{2\eta} E(t)^{2\eta} \left( \begin{array}{ccc} 1 & 0 \\ 0 & 1 \end{array} \right) = \mu^{2\eta} E(t)^{2\eta} \textbf{1}  \label{evenW}
\end{eqnarray}
In order to emphasize this point, we will denote even powers of the
perturbation operator $\hat W(t)$ without an operator head as $ W(t)^{2\eta}$ in the following. According to
(\ref{evenW}) for odd powers of the perturbation operator $\hat W(t)$ we have:
\begin{eqnarray}
\hat W(t)^{2\eta-1} &=& W(t)^{2(\eta-1)} \hat W(t) \\ 
\Longrightarrow [\hat W(t)^{2\eta-1}, \hat H_0(R)] &=& W(t)^{2(\eta-1)}[\hat W(t),\hat H_0 (R)]
\end{eqnarray}
Thus odd powers written in the form $\hat W^{2\eta -1}$ are proportional to the perturbation \textit{operator} $\hat W(t)$. 
As initial condition we fix \cite{Renziehausen09JPA, Renziehausen09Dip}:
\begin{eqnarray}
\vec \Psi(R, t=t_0) &=& \left( \begin{array}{c} \Psi_{1}(R,t=t_0) \\ 0 \end{array} \right) \label{startphi}
\end{eqnarray}
In the following calculations, for clarity, we suppress in the notation dependencies on the vibrational coordinate $R$ for all quantities. 
\subsection{Perturbation theory} \noindent
The starting point of time-dependent perturbation theory is the integral equation 
for the wave function \cite{Merzbacher98},
\begin{eqnarray}
\vec \Psi(t) = e^{-i \hat H_0 t } \vec \Psi (0) -i \int\limits^t_{0} dt' 
e^{-i \hat H_0 (t-t')} \hat W(t') \vec \Psi(t') \label{eight}  
\end{eqnarray}
where we assumed that the interaction starts at time $t_0=0$. In perturbation theory the wave function $\vec \Psi(t)$ is expanded in orders of the interaction-operator, and an approximative wave function $\vec \Psi(t,k)$ is obtained which contains all terms up to order $k$. 

The wave function in first order $\vec \Psi(t,1)$ is obtained
by substituting the exact wave function  $\vec \Psi(t')$ appearing under the integral by the
initial function (0$^{\textnormal{th}}$-order wave function) 
evolving in time with the system propagator ($\vec \Psi (t,0) = e^{-i \hat H_0 t } \vec \Psi (0)$):
\begin{eqnarray}
\vec \Psi(t,1) &=& e^{-i \hat H_0 t} \;\vec \Psi (0) -i \int\limits^t_{0} dt' 
\;e^{-i \hat H_0 (t-t')} \; \hat W(t') \; \vec \Psi(t',0) \label{nine} 
\end{eqnarray}
By iterating  (\ref{nine}) we obtain higher-order corrections as:
\begin{eqnarray}
\vec \Psi(t,k) &=& e^{-i \hat H_0 t} \; \vec \Psi (0) -i \int\limits^t_{0} dt'
\;e^{-i \hat H_0 (t-t')} \; \hat W(t') \; \vec \Psi(t',k-1) \label{nine-2}
\end{eqnarray}
We can use  (\ref{nine-2}) as a basis to devise a numerical algorithm for the calculation of 
perturbative wave functions \cite{Engel91CPC}. For this aim we  discretise the time $t$ into time steps $\Delta t$ yielding a time-grid where the times $t_n$ are defined as  $t_n = n\;\Delta t$ with whole-number values for $n$.
As a next step we set up an iteration scheme, where $\vec \Psi(t_{n+1},k)$ is calculated from $\vec \Psi(t_n,k)$. \newline
Therefore the integral in  (\ref{nine-2}) is divided into a first integral in the limits from $t'=0$ to $t'=t_n$, 
and a second one reaching from $t'=t_n$ and $t'=t_n+\Delta t$, yielding:
\begin{eqnarray}
\vec \Psi(t_{n+1},k) &=& e^{-i \hat H_0 \Delta t}\; \vec \Psi (t_n,k) \nonumber \\ 
&& - i 
\int\limits^{t_n + \Delta t}_{t_n} dt' \;e^{-i \hat H_0 (t_n + \Delta t -t')} \; 
\hat W(t') \; \vec \Psi(t',k-1) \label{eleven-1}  
\end{eqnarray}
This expression (\ref{eleven-1}) for the wave function can be interpreted easily: The first
term represents an unperturbed time-evolution of the system (during the time-interval
$\left [t_n,t_n + \Delta t \right ]$), whereas the second term stands for the possibility that during this interval at least one interaction takes place.\footnote{Here we note that of Ref. \cite{ Renziehausen09JPA}, Eqn. (9), which corresponds to  (\ref{eleven-1}) in this paper, is misleading, because there interactions taking place in the small time interval $\left [t_n, t' \right]$ were discarted.}
With these considerations a numerical scheme for the evaluation
of the wave functions can be developed, which we call the {\it simple algorithm} ($S$).
This algorithm is constructed by a replacement of the integral in  (\ref{eleven-1})
by a single term at the time $t'=t_{n+1}$, what leads to wave functions $\vec \Psi_S(t_n,\Delta t;k)$ depending on $t_n$, $\Delta t$ and $k$. Introducing the abbreviating notations $\vec \Psi (n,k) := \Psi_S(t_n,\Delta t;k)$  and $\hat W(n):=\hat W(t_n)$,
we get the following equation for the simple algorithm \cite{Renziehausen09JPA}: 
\begin{eqnarray}
\hspace{ -1 cm} \vec \Psi_S(n+1,k) &=& e^{-i \hat H_0 \Delta t}\; \vec \Psi_S (n,k) -i  
\Delta t  \; \hat W(n+1) \; \vec \Psi_S(n+1,k-1) \label{twelve}
\end{eqnarray}
with the start conditions $\vec \Psi_S(0,k) = \vec \Psi(0)$. 

As a last purpose in this chapter we explicate how the short-time propagator $e^{-i \hat H_0 \Delta t}$, which appears in (\ref{twelve}) and moves the wave function $\vec \Psi_{S}(n,k)$
over a time step $\Delta t$, can be executed numerically: This is customarily 
done by the split operator method of Feit and Fleck \cite{Feit82JCOMPP}, where a grid for the spatial coordinate $R$ is used.  Being a one step method correct in second order in the time step $\Delta t$, the application of the split-operator method in the simple algorithm $S$ does not diminish the order in the time-step $\Delta t$, in which the simple algorithm is correct. The reason for this is as we will see in the following Sec. \ref{error_wave} that the simple algorithm is a one step method which applies perturbation theory correctly only in first order in $\Delta t$.
\section{\label{error_wave} Error analysis of the wave functions $\vec \Psi_{S}(n,k)$} 
In this analysis of the wave functions $\vec \Psi_{S}(n,k)$, first we state a closed form for them, and then we show that the simple algorithm is a one step method correct in first order in $\Delta t$. 

Before we start our analytic analysis of the wave functions $\vec \Psi_{S}(n,k)$, we have to introduce notations that are important for the following. \newline
First, we define the sequence of non-commuting operators $\hat A_j$ when we use the product symbol $\prod$:
\begin{eqnarray}
\prod_{j=0}^n \hat A_j := \hat A_0 \hat A_1\cdots \hat A_n 
\end{eqnarray}
In the subsequent calculations combinatorial arguments are important. In particular, we will analyse combinatorial 
problems, where we have to calculate sums over all possible combinations with repetition. In these combinations $m$ elements are taken out of a set that contains $n$ elements, and the sequence in which the elements are chosen has no relevance. We specify a particular combination with repetition by a vector $ \vec \nu^{(n,m)}$ that has $n$ components, which are natural numbers or zero. The $j$-th component $ \nu^{(n,m)}_j$ of such a vector equals the number of cases how often the $j$-th element is chosen in this particular combination. By definition this implies
\begin{eqnarray}
\sum_{j=1}^n  \nu^{(n,m)}_j = m
\end{eqnarray}
Moreover, we introduce the combinatorial sum symbol $\Sigma_{\mathcal P_ {\vec \nu^{(n,m)}}}$,
\begin{eqnarray}
\sum_{\mathcal P_ {\vec \nu^{(n,m)}}}  f\left(\nu^{(n,m)}_1, \nu^{(n,m)}_2, \cdots , \nu^{(n,m)}_n\right) 
\end{eqnarray}
where $f$ is a function in the components of the vector $\vec \nu^{(n,m)}$, and where the sum contains all possible combinations with repetition for the situation that $m$ elements are taken out of a set with $n$ elements. E.g. it is obvious that for $n=2, m=3$ the equation
\begin{eqnarray}
\sum_{\mathcal P_ {\vec \nu^{(2,3)}}}  f \left (\nu^{(2,3)}_1, \nu^{(2,3)}_2 \right ) = f(3,0)+f(2,1)+f(1,2)+f(0,3)
\end{eqnarray}
is valid. Now the preparation for the virtual analysis is complete, so we can introduce
the announced closed form of the wave functions $\vec \Psi_{S}(n,k)$ 
\begin{eqnarray}
\hspace{-1.5 cm} \vec \Psi_{S}(n,k) = \left[ \sum_{m=0}^k (-i\; \Delta t)^m \sum_{\mathcal P_ {\vec \nu^{(n,m)}}} \prod_{j=0}^{n-1} \left( \hat W(n-j)^{\nu_{n-j}^{(n,m)}} e^{-i \hat H_0 \Delta t} \right) \right] \vec \Psi(0) \label{wavefunctionS}
\end{eqnarray}
This is proven in Appendix A. The expression (\ref{wavefunctionS}) allows an analysis of the norm given in Sec. \ref{error_norm}. 
\newline \newline
Now we show that the simple algorithm is a one step method that applies perturbation theory correctly in first order in $\Delta t$ for all $k \in \mathbb{N}$. Therefore we expand the with perturbation theory calculated wave function $\vec \Psi( \Delta t, k)$ for all $k \in \mathbb{N}$ in second order in the time step $\Delta t$, where we regard as a starting point the Eqn. (\ref{nine-2}) for $t= \Delta t$:
\begin{eqnarray}
\hspace{-2.5 cm} \vec \Psi(\Delta t,k) &=& e^{-i \hat H_0 \Delta t } \vec \Psi (0) -i \int\limits^{\Delta t}_{0} dt' 
e^{-i \hat H_0 (\Delta t-t')} \hat W(t') \vec \Psi(t',k-1) \nonumber \\
\hspace{-2.5 cm}  &=& \left [ 1-i \left ( \hat H_0 + \hat W(0) \right) - \frac{\Delta t^2}{2} \left ( \hat H_0^2 + \hat W(0) \hat H_0  \right. \right.  \nonumber \\
\hspace{-2.5 cm} && \left. \left. + \; \hat H_0 \hat W(0) + \left (1 - \delta_{k1} \right) W(0)^2 + i \left. \frac{\partial W(t')}{\partial t'}\right \vert_{t'=0} \right ) \right] \vec \Psi(0) +  \; \mathcal{O}(\Delta t^3) \label{exactexpand}
\end{eqnarray}
Then we calculate with Eqn. (\ref{twelve}) the wave function $\vec \Psi_S(1,k)$ for all $k \in \mathbb{N}$ by propagation of the start wave function $\vec \Psi(0)$ over one time step with the simple algorithm and expand the result in second order in $\Delta t$. For this calculation it is practicable to write $\hat W(\Delta t)$ instead of $\hat W(1)$:
\begin{eqnarray}
\hspace{-2.5 cm}  \vec \Psi_S (1, k) &=& e^{-i \hat H_0 \Delta t} \vec \Psi(0) - i \Delta t \hat W(\Delta t) \vec \Psi_S(1, k-1) \nonumber \\
\hspace{-2.5 cm}  &=& e^{-i \hat H_0 \Delta t} \vec \Psi(0)  - i \Delta t \hat W (\Delta t) e^{-i \hat H_0 \Delta t} \; \times  \nonumber \\
\hspace{-2.5 cm}  && \times \; \left [ e^{-i \hat H_0 \Delta t} \vec \Psi(0) - i \Delta t \left ( 1- \delta_{k1} \right) \hat W(\Delta t) \vec \Psi_S(1, k-2) \right] \nonumber \\
\hspace{-2.5 cm}  &=&  \left [e^{-i \hat H_0 \Delta t} - i \Delta t \hat W(\Delta t) e^{-i \hat H_0 \Delta t} - \Delta t^2 \left( 1 - \delta_{k1} \right) W(\Delta t)^2 e^{-i \hat H_0 \Delta t} \right ] \vec \Psi(0)  \nonumber \\
\hspace{-2.5 cm}  && +  \; \mathcal{O}(\Delta t^3) \nonumber \\
\hspace{-2.5 cm}  &=& \left [1 - i \left ( \hat H_0 + \hat W(0) \right) \Delta t - \frac{\Delta t^2}{2} \left ( \hat H_0^2 + 2 \hat W(0) \hat H_0  \right. \right. \nonumber \\
\hspace{-2.5 cm} && \left. \left. + \; 2 \left (1-\delta_{k1} \right) W(0)^2 + 2i \left. \frac{\partial W(t')}{\partial t'}\right \vert_{t'=0} \right) \right] \vec \Psi(0)  +  \; \mathcal{O}(\Delta t^3) \label{simpleexpand}
\end{eqnarray}
By comparison of (\ref{exactexpand}) and (\ref{simpleexpand}) it can be cognized that
\begin{eqnarray}
\hspace{-2.0 cm}  \vec \Psi(\Delta t,k) - \vec \Psi_S (1, k) &=& \frac{\Delta t^2}{2}  \left ( \hat W(0) \hat H_0 - \hat H_0 \hat W(0) \right. \nonumber \\
\hspace{-2.0 cm}  && \left. + \;  \left (1-\delta_{k1} \right ) W(0)^2  + i \left. \frac{\partial W(t')}{\partial t'}\right \vert_{t'=0}     \right) \vec \Psi(0)  +  \; \mathcal{O}(\Delta t^3) \label{differror}
\end{eqnarray}
so the simple algorithm is a one step method which applies perturbation theory correctly in first order in $\Delta t$ for all $k \in \mathbb{N}$. Due to standard numerical textbook analysis of the asymptotic development for the global discretisation error of one step methods \cite{Stoer80}, the difference between the with perturbation theory calculated wavefunction $\vec \Psi(t,k)$ and the wave function propagated over the time $t$ with the simple algorithm  $\vec \Psi_S(n,k)$, with $n=t/\Delta t$, is given by\footnote{The associated proposition in \cite{Stoer80} to the Eqn.(\ref{globalerror}) was discussed there for real functions instead of a complex wave function but it is straightforward to see that this is no limitation for an application of this proposition here. Futhermore we suppose that the electric field $E(t)$ and therefore the wave function $\vec \Psi(t)$, too, depends in practical applications of the simple algorithm smoothly on time so that the proposition requirements given in \cite{Stoer80} are not violated.}: 
\begin{eqnarray}
\vec \Psi(t,k) - \vec \Psi_S(n,k) = \Delta t \;  \vec \chi(t,k) + \; \mathcal{O}(\Delta t^2) \label{globalerror}
\end{eqnarray}
For the function $\vec \chi(t,k)$ appearing in the above equation holds that it is independent of the time step $\Delta t$ and it fulfils $\vec \chi(0,k) = 0$. 

For the evaluation of the accuracy of the simple algorithm has to be taken into account that the with perturbation theory calculated wave function $\vec \Psi (t,k)$ is itself an approximative solution of the Schrödinger equation (\ref{tdse}), which deviates from the exact solution $\vec \Psi (t)$. This deviation can be noted as $\vec \Psi(t) - \vec \Psi(t,k) = \vec \phi(t,k)$. Therefore for the difference between $\vec \Psi_S(n,k)$ and the exact solution $\vec \Psi(t)$ holds:
\begin{eqnarray}
\vec \Psi(t) - \vec \Psi_s(n,k) = \vec \phi(t,k) + \Delta t \;  \vec \chi(t,k) + \; \mathcal{O}(\Delta t^2) \label{globalerror2}
\end{eqnarray}
As the main conclusion of this section, the simple algorithm is a one step method that applies perturbation theory correctly for all $k \in \mathbb{N}$ only in first order in $\Delta t$, so one might think that it is not useful to calculate wave functions $\vec \Psi_S(n,k)$ with $k>1$. However, this reasoning is not correct because the use of a higher perturbation order $k$ takes according to (\ref{globalerror2}) influence on the difference $\vec \Psi(t)-\vec \Psi_S(n,k)$ and stabilizes so the simple algorithm against divergences of the norm, see the appendant discussions in Sec. \ref{Interpretation}.

\section{\label{error_norm} Error analysis of the norms for the wave function $\vec \Psi_{S}(n,k)$} 
\subsection{\label{error_pre} Preliminary remarks}
From the closed form expression (\ref{wavefunctionS}) for the wave function $\vec \Psi_{S}(n,k)$, it emerges that the wave function can be decomposed into terms $\vec \Psi_{m,S}(n)$ for different orders of perturbation: 
\begin{eqnarray}
\vec \Psi_S(n,k) &=& \sum_{m=0}^k \vec \Psi_{m,S}(n)  \label{orderfunc} 
\end{eqnarray}
with 
\begin{eqnarray} \hspace{-2,0cm}  \vec \Psi_{m,S}(n)  =
(-i\; \Delta t)^m \sum_{\mathcal P_ {\vec \nu^{(n,m)}}} \prod_{j=0}^{n-1} \left( \hat W(n-j)^{\nu_{n-j}^{(n,m)}} e^{-i \hat H_0 \Delta t} \right) \vec \Psi(0) \label{orderfunc2}
\end{eqnarray}
where the parameter $m$ denotes the order of perturbation. The wave functions $\vec \Psi_{m,S}(n)$ have a clear interpretation as they are related to an interaction between the laser pulse and the molecular system, where $m$ photons are exchanged, and therefore the electronic state changes $m$ times during the time interval $[0,n \hspace{0.07cm} \Delta t]$. 
Employing the decomposition (\ref{orderfunc}), the norm yields:
\begin{eqnarray}
N_{n,S}^k &=& \left \langle \vec \Psi_S(n,k)\vert \vec \Psi_S(n,k) \right \rangle  \nonumber \\ 
          &=& \sum_{j=0}^k \sum_{h=0}^k \left \langle \vec \Psi_{j,S}(n)\vert \vec \Psi_{h,S}(n) \right \rangle \label{normexpand} 
\end{eqnarray}
Using the substitution $p=j+h$, we can transform (\ref{normexpand}) regarding that the norm of the initial wave function $\vec \Psi(0)$ is defined as  $\left \langle \vec \Psi(0) \vert \vec \Psi (0) \right \rangle = 1$ as follows:
\begin{eqnarray}
N_{n,S}^k &=& \sum_{j=0}^{k} \sum_{p=j}^{j+k}  \left \langle \vec \Psi_{j,S}(n)\vert \vec \Psi_{p-j,S}(n) \right \rangle \nonumber \\ 
          &=& \sum_{p=0}^{2k} \sum_{j= \max(0,p-k)}^{\min(p,k)} \left \langle \vec \Psi_{j,S}(n)\vert \vec \Psi_{p-j,S}(n) \right \rangle \nonumber \\
          &=& 1 + \sum_{p=1}^{2k} \sum_{j= \max(0,p-k)}^{\min(p,k)} \left \langle \vec \Psi_{j,S}(n)\vert \vec \Psi_{p-j,S}(n) \right \rangle                \label{normexpand2} 
\end{eqnarray} 
We note that terms for odd $p$ in  (\ref{normexpand2}) are zero, which results from the choice of the initial conditions, where only the electronic state $|1\ra$ is populated (see  (\ref{startphi})).  These terms involve the scalar product of wave functions in the different electronic states $|0\ra, |1\ra$, which are orthogonal. Due to this connection we substitute $p=2m$ in  (\ref{normexpand2}), which yields:
\begin{eqnarray}
N_{n,S}^k &=& 1 + \sum_{m=1}^{k} \sum_{j= \max(0,2m-k)}^{\min(2m,k)} \left \langle \vec \Psi_{j,S}(n)\vert \vec \Psi_{2m-j,S}(n) \right \rangle \nonumber \\
&:=&  1 +  \sum_{m=1}^{k} N_{n,2m,S}^k, \label{normexpand3} 
\end{eqnarray}
where the terms $N_{n,2m,S}^k$, given by
\begin{eqnarray}
N_{n,2m, S}^k &=& \sum_{j= \max(0,2m-k)}^{\min(2m,k)} \left \langle \vec \Psi_{j,S}(n)\vert \vec \Psi_{2m-j,S}(n) \right \rangle, \label{normexpand4} 
\end{eqnarray}
characterize norm deviations from unity. As was already discussed in \cite{Renziehausen09JPA} and \cite{Renziehausen09Dip} these orders $N_{n,2m,S}^k$ can be decomposed into two different types: For $2k \geq  2m > k$ we call these orders \textit{oscillatory orders}:  
\begin{eqnarray} 
N_{n,2m, S}^k &=& \sum_{j = 2m-k}^k \left \langle \vec \Psi_{j,S}(n)\vert \vec \Psi_{2m-j,S}(n) \right \rangle, \label{normexpand5} 
\end{eqnarray}
and for $k\geq 2m >0 $ they are called \textit{stationary}: 
\begin{eqnarray} 
N_{n,2m, S}^k &=& \sum_{j = 0}^{2m} \left \langle \vec \Psi_{j,S}(n)\vert \vec \Psi_{2m-j,S}(n) \right \rangle. \label{normexpand6} 
\end{eqnarray}
Note that the stationary orders contain no explicit dependence on $k$ any more. \newline
In \cite{Renziehausen09JPA}, it has been shown that these two kinds of orders behave differently when parameters in the numerical simulation are changed and thus they have different physical interpretations. The former presented discussion was based on equations, which were presented without proof. In what follows the missing derivation will be given for the simple algorithm.
\subsection{\label{error_norm_D} Norm analysis for the simple algorithm \textit{S}} 
In the norm analysis for the simple algorithm in this chapter, we state first that the stationary orders can be written in a closed form. Secondly we discuss that for the oscillatory norm orders we can introduce an approximation that allows to analyse how these norm orders scale in the time step $\Delta t$, and in this context we explain an evidence called annihilation thesis, which is a pre-condition for the approximation. Thirdly we show that we can easily apply this approximation method, implemented foremost for the oscillatory orders, for the stationary orders, too. \newline
In the calculations given here we are focussing on the considerations referring to the mentioned approximation because the most important point for the understanding of the mathematical background of the results presented in \cite{Renziehausen09JPA} is to get the idea how this approximation is introduced. Other, more straightforward calculations are given in the appendices. \newline \newline
Thus, for the stationary norm orders $N_{n,2m, S}^k$ of the perturbative wave functions calculated via the simple algorithm we first note that
\begin{eqnarray}
N_{n,2m, S}^k = (-1)^m \Delta t^{2m} \sum_{\mathcal P_{\vec \nu^{(n,m)}}} \prod_{j=1}^n W(j)^{2 \; \nu_j^{(n,m)}}; \label{statSRule}
\end{eqnarray}
the proof of this equation is given in \ref{appendix C}. We emphasize that in  (\ref{statSRule}) only even orders of $\hat W(n)$ appear, for which (\ref{evenW}) is valid. 
\newline \newline
After having presented the closed form (\ref{statSRule}) for the stationary orders of the simple algorithm $N_{n,2m, S}^k, k \geq 2m > 0$, we will investigate now the announced properties of the oscillatory orders $N_{n,2m,S}^k, k< 2m \leq 2k$. To do so, we employ  (\ref{normexpand5}) to calculate that for the orders $N_{n,2m,S}^k, k< 2m \leq 2k$ the following equation holds:
\begin{eqnarray}
\hspace{-2,5cm} N_{n,2m, S}^k  &=& \sum_{j = 2m-k}^{k} \left \langle \vec \Psi_{j,S}(n) \vert \vec \Psi_{2m-j,S}(n) \right \rangle \nonumber \\ 
\hspace{-2,5cm}                &=&\Delta t^{2m} \hspace{-0.3cm} \sum_{j = 2m-k}^{k} \sum_{\mathcal P_{\vec \nu^{(n,j)}}} \sum_{\mathcal P_{\vec \rho^{(n,2m-j)}}} (-1)^{m-j}  \left \langle \vec \Psi(0) \left  \vert \prod_{q=1}^{n} \left ( e^{ i \hat H_0 \Delta t} \hat W(q)^{\nu_q^{(n,j)} } \right ) \right. \right.  \hspace{-0.1cm} \times \nonumber \\
\hspace{-2,5cm}                && \left. \left. \times \; \prod_{p=0}^{n-1} \left ( \hat W(n-p)^{\rho_{n-p}^{(n,2m-j)}} e^{ -i \hat H_0 \Delta t } \right) \right \vert \vec \Psi(0) \right \rangle \label{prostatS6.1}
\end{eqnarray}
Each summand of the sums in  (\ref{prostatS6.1}) over $j, \mathcal P_{\vec \nu^{(n,j)}}$ and $\mathcal P_{\vec \rho^{(n,2m-j)}}$ is a product of the sign factor $(-1)^{m-j}$ and a bracket term $\left \langle \vec \Psi(0) \left \vert \cdots \right \vert \vec \Psi(0) \right \rangle$. The same bracket terms $\left \langle \vec \Psi(0) \left \vert \cdots \right \vert \vec \Psi(0) \right \rangle$ appear multiple in different summands of these sums, but the  $j$-value of these summands differs, and thus they have different sign factors $(-1)^{m-j}$. Due to these different sign factors, these summands must cancel (at least) partly each other, and for each of the various types of bracket terms in the sums in (\ref{prostatS6.1}) can only survive bracket terms with either the sign factor $(-1)$ or $(+1)$. Now it can be shown that all these surving bracket terms do \textit{not} have \textit{different} signs but the \textit{same} sign, scilicet $(-1)^{k-m}$. This coherence is from now on called the \textit{annihilation thesis} and the proof of it is presented in \ref{appendix D}. \newline
Our aim is to count the total number $ \#_{n,2m,S}^k $ of all the surviving bracket terms in  (\ref{prostatS6.1}) because the scaling of the number of these terms in the time step $\Delta t$ helps to draw conclusions how the oscillatory order $N_{n,2m, S}^k$ scales in the time step $\Delta t$. Employing the annihilation thesis allows to count the total number of all the surviving bracket terms, and it can be concluded that:
\begin{eqnarray}
\#_{n,2m,S}^k = \left \vert \sum_{j = 2m-k}^{k} \left [ (-1)^{j} \left ( \sum_{\mathcal P_{\vec \nu^{(n,j)}}} 1  \right ) \left ( \sum_{\mathcal P_{\vec \rho^{(n,2m-j)}}} 1 \right )  \right ] \right \vert \label{nummer1}
\end{eqnarray}  
Now we introduce an approximation for the calculation of the norm orders $N_{n,2m,S}^k$ by approximating each of the surviving summands in  (\ref{prostatS6.1}) with a constant factor $\overline{W}^{2m}$ and we take into account the fact that all the surviving terms have the sign-prefactor $(-1)^{k-m}$ by a global sign prefactor $(-1)^{k-m}$ for the calculation of the norm orders  $N_{n,2m,S}^k$: 
\begin{eqnarray}
\hspace{-2,0cm} N_{n,2m,S}^k &=& (-1)^{k-m} \Delta t^{2m} \overline{W}^{2m} \#_{n,2m,S}^k \nonumber \\ 
\hspace{-2,0cm} &=& (-1)^{k-m} \Delta t^{2m} \overline{W}^{2m}  \left \vert \sum_{j = 2m-k}^{k} \left [ (-1)^{j} \left ( \sum_{\mathcal P_{\vec \nu^{(n,j)}}} 1  \right ) \left ( \sum_{\mathcal P_{\vec \rho^{(n,2m-j)}}} 1 \right )  \right ] \right \vert  \label{prostat8}
\end{eqnarray}
Since the sum $\sum_{\mathcal P_{\vec \nu^{(n,j)}}}$ is executed over all combinations with repetition for a choice of $j$ elements out of a set with $n$ elements, elementary statistics yield \cite{Makinson08}:
\begin{eqnarray}
 \sum_{\mathcal P_{\vec \nu^{(n,j)}}} 1 &=&  { n+j-1 \choose j} = \frac{(n+j-1)(n+j-2) \cdots (n+1) n}{j!} \nonumber \\
                                       &=& \frac{n^j}{j!} + \left ( \sum_{q=1}^{j-1}  q \right ) \frac{n^{j-1}}{j!} + \mathcal{O} \left[ \left ( \frac{1}{n} \right)^{ 2-j } \right] \nonumber \\
                                       &=&   \frac{n^j}{j!} \left \lbrace 1+ \frac{1}{2} \frac {j (j-1)} {n} + \mathcal{O} \left [  \left( \frac{1}{n} \right )^2 \right] \right \rbrace \label{prostat9}
\end{eqnarray}
Inserting (\ref{prostat9}) in  (\ref{prostat8}) and shifting the sum index $j$ by $m$, we find
\begin{eqnarray}
\hspace{-2,5cm} N_{n,2m,S}^k &=& (-1)^{k-m} \Delta t^{2m} \overline{W}^{2m} n^{2m} \left \vert \sum_{j = -(k-m)}^{k-m} \frac{(-1)^j}{(m+j)!(m-j)!} \right \vert \left[ 1 + \mathcal{O} \left( \frac{1}{n} \right) \right]  \nonumber \\
\hspace{-2,5cm}              &:=&  (-1)^{k-m} \Delta t^{2m} \overline{W}^{2m} n^{2m} \left \vert \xi(m,k) \right \vert \left[ 1 + \mathcal{O} \left( \frac{1}{n} \right) \right], \label{prostat10}
\end{eqnarray}
where we introduced the help function $\xi(m,k)$:
\begin{eqnarray}
\xi(m,k) &=&  \sum_{j = -(k-m)}^{k-m} \frac{(-1)^j}{(m+j)!(m-j)!} \label{helpeqnextra} 
\end{eqnarray}
With an induction proof over $k$ one can show that if the oscillatory condition $k < 2m \leq 2k$ is valid, $\xi(m,k)$ can be written in the form:
\begin{eqnarray}
\xi(m,k) &=& \frac { (-1)^{k-m} }{m}  \frac{ 1 }{k! (2m-1-k)!} \label{helpeqn} 
\end{eqnarray}
As the base case for this induction proof we choose $k=m$, because for the validity of the oscillatory condition $k < 2m \leq 2k$ this is the lowest possible value for $k$, and we calculate with  (\ref{helpeqnextra}) that $\xi(m,m)=1/(m!)^2$, which fits with the result we get for $k=m$ with  (\ref{helpeqn}). Then we perform the induction step assuming as induction hypothesis that for an arbitrarily chosen value for $k$ satisfying the oscillatory condition (\ref{helpeqn}) is true:
\begin{eqnarray}
\hspace{-2,0cm} \xi(m,k+1) &=& \sum_{j = -(k+1-m)}^{k+1-m} \frac{(-1)^j}{(m+j)!(m-j)!} \nonumber \\
\hspace{-2,0cm}           &=& (-1)^{k+1-m} \frac{2}{(k+1)!(2m-k-1)!} + \sum_{j=-(k-m)}^{k-m}  \frac{(-1)^j}{(m+j)!(m-j)!} \nonumber \\
\hspace{-2,0cm}           &=& (-1)^{k+1-m} \frac{2}{(k+1)!(2m-k-1)!} + \frac { (-1)^{k-m} }{m} \frac{ 1 }{k! (2m-1-k)!} \nonumber \\
\hspace{-2,0cm}           &=&  \frac { (-1)^{k+1-m} }{m} \frac{ 1 }{(k+1)! \left[2m-1-(k+1)\right]!} \label{prostat12} 
\end{eqnarray}
This is (\ref{helpeqn}) for the shift $k \rightarrow k+1$, so the induction step and the proof of  (\ref{helpeqn}) is complete. \newline
For a re-writing of  (\ref{prostat10}), we first substitute the expression (\ref{helpeqn}) for  $\xi(m,k)$ in (\ref{prostat10}), and second, introducing the propagation time $t=n \; \Delta t$, we substitute furthermore $\frac{t}{\Delta t}$ for $n$ in (\ref{prostat10}). So as a final result we get Eqn. (14) of Ref. \cite{Renziehausen09JPA} for the simple algorithm:
\begin{equation}
N_{t,2m,S}^k = \frac{ (-1)^{k-m}}{m} \frac{ t^{2m}} {k! (2m-1-k)!} \overline{W}^{2m} + \mathcal{O}(\Delta t) \label{14Sold}
\end{equation}
In the latter equation we have replaced the index $n$ on the left side by $t$ referring explicitly to this time. \newline \newline
As the third task in this chapter, the above approximation can be applied not only for the oscillatory orders $ N_{n,2m,S}^k, k < 2m \leq 2k$, but as well for the stationary orders $ N_{n,2m,S}^k, k\geq 2m >0 $, because in (\ref{statSRule}) it can be seen that all terms appearing in the sum over $\mathcal P_{\vec \nu^{(n,m)}}$ have the same global sign $(-1)^m$.  
Therefore we approximate all factors $W(j)^2$ by $\overline{W}^2$ in  (\ref{statSRule}). Regarding  (\ref{prostat9}), this leads to
\begin{eqnarray} 
N_{n,2m,S}^k &=& (-1)^m \Delta t^{2m} \sum_{\mathcal P_{\vec \nu^{(n,m)}}} \prod_{j=1}^n W(j)^{2 \; \nu_j^{(n,m)}} \nonumber \\
             &=& (-1)^m \Delta t^{2m} \overline{W}^{2m} \sum_{\mathcal P_{\vec \nu^{(n,m)}}} 1 \nonumber \\
             &=& (-1)^m \Delta t^{2m} \overline{W}^{2m} \frac{n^m}{m!}  \left [ 1 + \mathcal{O} \left (\frac{1}{n} \right ) \right] 
\end{eqnarray}
Substituting $ n = \frac{t}{\Delta t} $ we receive the final result
\begin{eqnarray}
N_{t,2m,S}^k = (-1)^m \frac{t^m}{m!} \Delta t^m \overline{W}^{2m} + \mathcal{O} \left( \Delta t^{m+1} \right), \label{13Sold} 
\end{eqnarray}
which is the Eqn. (13) in  Ref. \cite{Renziehausen09JPA} for $G=S$. \newline \newline
In Sec. \ref{Interpretation} we will discuss the consequences of the results derived in this chapter.
\section{\label{Interpretation} Interpretation} \noindent
\setlength{\unitlength}{1.0cm} \begin{figure} 
\resizebox{0.8 \textwidth}{!}{\includegraphics[width=13cm,height=8.5cm]{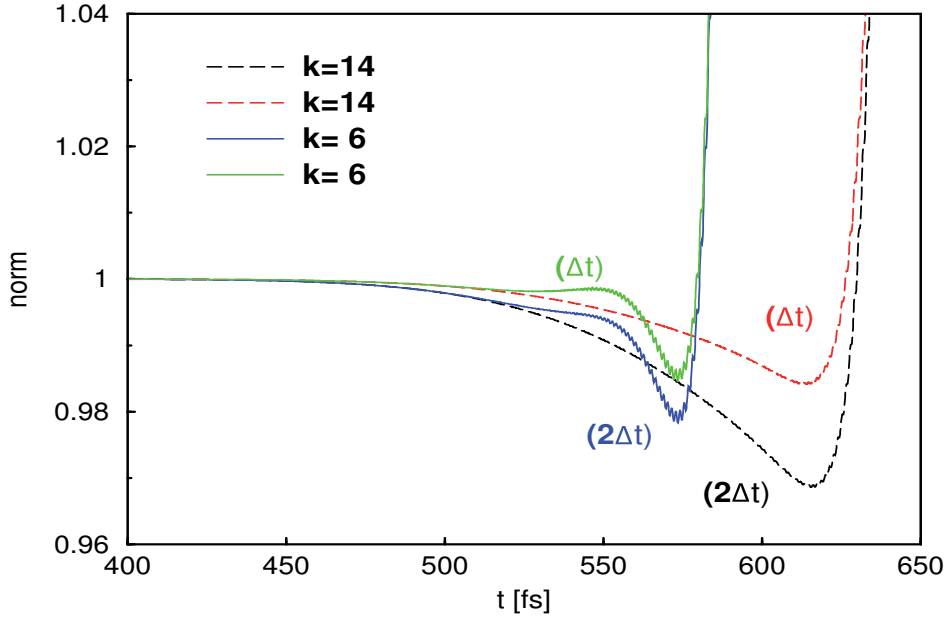}} 
\caption{
Time-dependent norm of wave functions determined by perturbation theory
(simple algorithm) of different orders $k$, as indicated.
Curves obtained in different order $k$ of perturbation theory are
compared for time steps of $\Delta t$ = 4 $\times 10^{-2}$ fs and
$\Delta t' = 2 \Delta t = 8 \times 10^{-2}$ fs.  
}
\label{fig3alt}
\end{figure}
In this chapter we will pick up the areas of interest established in Sec. 2.3 and Sec. 3 of \cite{Renziehausen09JPA}, explain the mathematical background to the discussions related to the simple algorithm we started there and carry them forward.  In line with this task we give the results derived in Sec. \ref{error_norm} a physical interpretation. Moreover, we present in a gedankenexperiment an analogy of perturbation theory to a scattering experiment. 

As a starting point we will discuss the stationary orders. These orders depend in leading order on the $m$-th power of the time step $\Delta t$. For a not too long propagation time $t$ and a small time step $\Delta t$, the stationary order for $m=1$ (calculated with  (\ref{13Sold}))
\begin{eqnarray}
N_{t,2,S}^k = - t \; \Delta t \; \overline{W}^2 + \mathcal{O}(\Delta t^2) \label{interpret1}
\end{eqnarray}
gives the dominant contribution to the norm deviation. Thus, the norm deviations caused by the stationary orders are negative, and they are independent of the order $k$ but they depend approximately linearly on the time step $\Delta t$. So, in the limit $\Delta t \rightarrow 0$, the norm deviations caused by the stationary orders vanish. Thus the stationary orders are related to the $\Delta t$-dependent part in the difference of the approximative wave function $\vec \Psi_S(n,k)$ to the exact wave function $\vec \Psi(t)$ (see Eqn. (\ref{globalerror2})).  From these findings we conclude that these errors are numerical errors which arise using the simple algorithm from the approximation of the integral in  (\ref{eleven-1}) by only one summand, and they have no physical meaning. This interrelation was discussed in the explanations to figure 3 in \cite{Renziehausen09JPA}, shown in this paper for clarity again in figure \ref{fig3alt}, 
where one can realize first that for early propagation times a bisection of the time step $\Delta t$ leads to a bisection of the norm deviations caused by the stationary orders, and second that these norm deviations are the same for equal time steps $\Delta t$ but different orders $k$.

\begin{figure}  
\resizebox{0.8 \textwidth}{!}{\includegraphics[width=13cm,height=8cm]{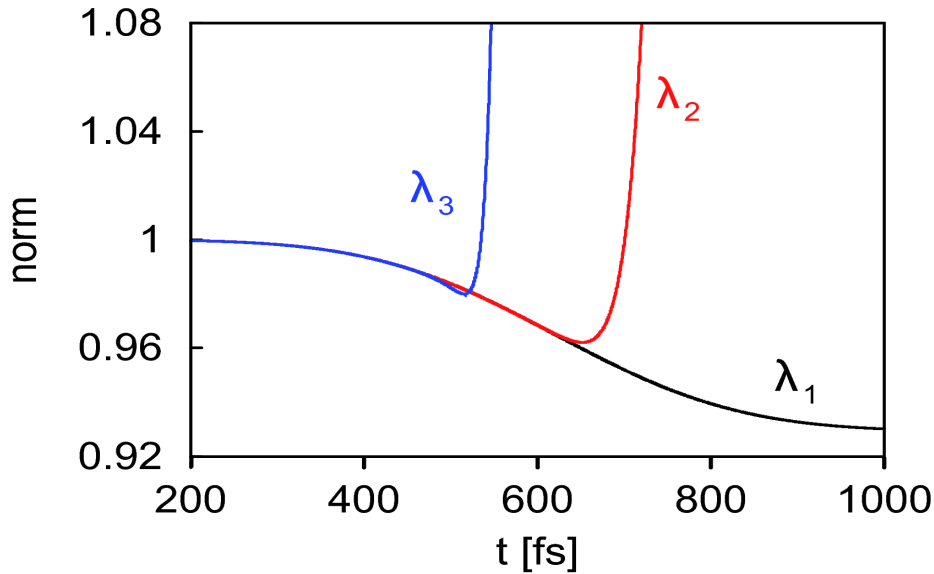}}
\caption{
Time-dependent norm calculated with the simple algorithm.
The calculations ($k$ = 14) involved different
potential gradients $m_0=\lambda_i \times 10^{-3}$ a.u., with numerical values of
$\lambda_1 =3, \lambda_2 =2 $ and $\lambda_3 = 1$.
Whereas for the steepest potential the norm is conserved, divergences occur
for smaller gradients where the wave-packet motion proceeds slower. 
}
\label{fig5alt}
\end{figure}

According to (\ref{evenW}) and (\ref{statSRule}) the stationary orders $N_{n,2m,S}^k$ only depend on the time step $\Delta t$, the dipole moment $\mu$ and the electric field $E(t)$, but they do not depend on the Hamilton operator $\hat H_0$ of the unperturbed system. This is one more evidence that the stationary orders are only numerical errors.
Thus in particular they do not depend on the shape of the potential surfaces $ V_j(R), j \in \lbrace 0,1 \rbrace$  of the two electronic states $ \vert 1 \rangle $ and  $ \vert 0 \rangle $, nor on the propagation of the wave function components $ \Psi_1(R,n)$ and $\Psi_0(R,n)$ (here we explicitly wrote out the $R$-dependence) onto these surfaces. This context was presented in the discussion of figure 5 of \cite{Renziehausen09JPA}, shown in this paper in figure \ref{fig5alt} for the case when the simple algorithm is used: For a simulation model with linear potentials 
\setlength{\unitlength}{1.0cm}  
\begin{equation}
 V_j(R) = m_j R + C_j, \; j \in \lbrace 0,1 \rbrace, \; m_1 = -m_0  \label{potgrad}
\end{equation}
we varied in \cite{Renziehausen09JPA} the parameter for the steepness of the potentials $m_0$. As a result of the simulations it is obvious that for small propagation times the negative norm deviations caused by the stationary orders are the same.

More explicitly, we compute the norm deviation caused by the stationary orders $N_{n,2m,S}^k, k \geq 2m > 0$ for an electric field given by
\begin{eqnarray}
E(t) = A(t) \cos \left (\Phi(t) \right ), \label{interpret2} 
\end{eqnarray}
where $ A(t) $ is the envelope of the electric field and $ \Phi(t) $ is a fast oscillating phase function. Since we can assume that the lowest stationary order $N_{n,2,S}^k$ gives the dominant contribution to the norm deviations, it is in our interest to calculate this order analytically. Therefore we consider that permutation vectors of the form $\vec \nu^{(n,1)}$ have $n$ components at which one component is 1 and all other components are 0. So we can simplify (\ref{statSRule}) for $m=1$:
\begin{eqnarray}
N_{n,2,S}^k &=& - \Delta t^2 \sum_{\mathcal P_{\vec \nu^{(n,1)}}} \prod_{j=1}^n W(j)^{2 \nu_j^{(n,1)}} \nonumber \\
            &=& - \Delta t \; \sum_{j=1}^n W(j)^2 \Delta t \label{interpret3} 
\end{eqnarray}
Inserting (\ref{evenW}) and (\ref{interpret2}) into (\ref{interpret3}) and approximating the sum by an integral leads to the result:
\begin{eqnarray}
N_{t,2,S}^k = - \mu^2 \; \Delta t \int_{0}^{t} dt' \; A(t')^2 \cos^2 \left (\Phi(t') \right )  \label{interpret4} 
\end{eqnarray}
Here, we denoted the time dependency again by $t$. Since this integral in the limits $0$ and $t$ is proportional to the stored energy in the laser pulse in this time interval, the norm deviations caused by the stationary orders at a certain point in time $t$ are approximately a proportion for this quantity. Therefore for the limit $ t \rightarrow \infty$, the stationary order $N_{t,2,S}^{k}$ is proportional to the total energy of the laser pulse. Within the ''slow-varying envelope approximation'' \cite{Meystre91,Scully97} we can approximate (\ref{interpret4}) by substituting the factor  $ \cos^2 \left (\Phi(t') \right ) $ under the integral by $\frac{1}{2}$ and get as a result:
\begin{eqnarray}
N_{t,2,S}^k = - \frac{ \mu^2 \; \Delta t }{2} \int_{0}^{t} dt' \; A(t')^2 \label{interpret5} 
\end{eqnarray}
This result reveals that if the ''slow-varying envelope approximation'' is valid, the norm deviations caused by the stationary orders do not depend approximately on the phase $\Phi(t)$ but on the time integral over the squared envelope of the electric field. 
Futhermore it can be cognized from (\ref{interpret5}) that $N_{t,2,S}^k$ depends linearly on the time step $\Delta t$. This result is related to the fact that according to Eqn. (\ref{globalerror2}) the leading order of the $\Delta t$-dependent part of the deviations of the wavefunctions $\vec \Psi_S(n,k) $ to the exact solution $\vec \Psi(t)$ is the first order.

In particular we regard as an example for the use of (\ref{interpret5}) the Gaussian laser pulse modified by a linear spectral chirp $b_2$ we employed for our numerical application of the simple algorithm in \cite{Renziehausen09JPA} and compare the results we got there numerically with the norm deviations we are now able to calculate analytically with (\ref{interpret5}): \newline  The unchirped pulse is given by
\begin{eqnarray}
E(t) =  E_0' e^{- \beta' \left (t-t_d \right )^2}  \cos \left [ \omega_0 \left ( t - t_d \right ) \right ]  \label{interpret6} 
\end{eqnarray}
with an envelope having a full width at half maximum of $ \tau' = \sqrt{4 \ln 2 / \beta'} $,  and the chirped laser pulse is given by  
\begin{eqnarray}
E(t) =  E_0 e^{- \beta \left (t-t_d \right )^2} \cos \left [ \omega_0 \left ( t - t_d \right ) + \frac{a_2}{2} \left (t-t_d \right )^2 \right ]  \label{interpret7} 
\end{eqnarray} 
In the last equations, the field strengths are denoted as $E_0'$ and $E_0$ for the unchirped and for the chirped fields, respectively, and $t_d$ denotes the point in time when the envelope of the field is maximal. The various parameters appearing in the equations for the shaped and for the unshaped electric fields are related as follows \cite{Wollenhaupt05springer}:
\begin{eqnarray} 
E_0  &=& \sqrt{\frac{1}{1 + 2 i  \beta' b_2}} E_0' ,\;\;\; 
\mid E_0 \mid \; = \left (1 + 4 \beta'^2 b_2^2 \right)^{-\nicefrac{1}{4}} \mid E_0' \mid \nonumber \\
\beta &=& \frac{1}{ \frac{1}{\beta'} + 4 \beta' b_2^2}  \label{interpret8} \\  
a_2 &=& \frac{b_2}{ \frac{1}{4 \beta'^2} + b_2^2}.  \nonumber  
\end{eqnarray}
\setlength{\unitlength}{1.0cm} \begin{figure} 
\resizebox{0.8 \textwidth}{!}{\includegraphics[
width=18cm,height=21cm
]{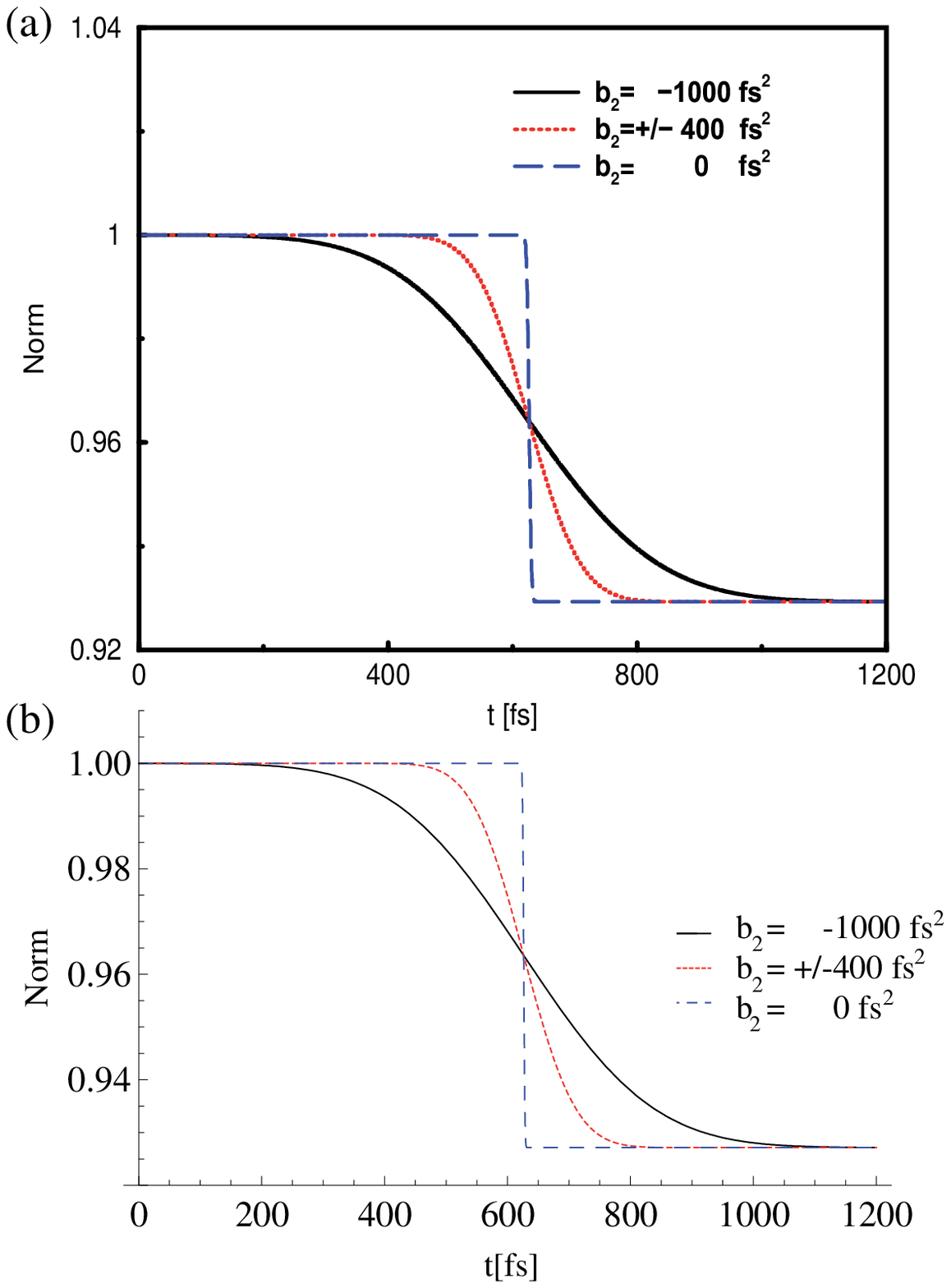}}
\caption{
The norm deviations arising from a numerical simulation employing the simple algorithm (panel (a)), and the norm deviations calculated analytically with  (\ref{interpret9}) (panel (b)). The analytically calculated deviations in panel (b) accurately match the numerical results in panel (a).
}
\label{fig9a}
\end{figure}
For the electric field given by (\ref{interpret7}) we can calculate with (\ref{interpret5}) and (\ref{interpret8}) that the norm deviations caused by the stationary orders are given by
\begin{eqnarray}
\hspace{-2,0cm} N_{t,2,S}^k = - \mu^2 \; \Delta t \; E_0'^2 \tau' \sqrt{ \frac{ \pi }{ 128 \ln 2}} \left \lbrace 1 + \textnormal{erf} \left [ \sqrt{ \frac{32 \ln 2 \; \tau'^2} { \tau'^4 + \left (16 \ln 2 \;  b_2 \right )^2} } \left ( t - t_d \right ) \right ] \right \rbrace. \label{interpret9}
\end{eqnarray}
Norm deviations calculated with (\ref{interpret9}) match excellently with the numerical results which are presented in figure 4 of \cite{Renziehausen09JPA}. This relation is presented in figure \ref{fig9a}, where the norm deviations from unity caused in the numerical simulation with the simple algorithm are shown again in panel (a), and the norm deviations from unity calculated analytically are shown in panel (b). For the analytical calculation we used the parameter values which lead to the numerical results presented in figure 4 of \cite{Renziehausen09JPA}, namely   $\mu = 1$ a.u., $\Delta t = 3.31$ a.u., $\tau' = 4.13*10^2$ a.u. and $E_0' = 1.19*10^{-2}$ a.u. and different values for the spectral chirp parameter $b_2$, which are given in figure \ref{fig9a}. 

Furthermore, in the limit $t \rightarrow \infty$, one obtains from   (\ref{interpret9})
\begin{eqnarray}
\lim_{t \rightarrow \infty} N_{t,2,S}^k = - \mu^2 \; \Delta t \; E_0'^2 \;  \tau' \sqrt{ \frac{\pi}{32 \ln 2} }. \label{interpret10}
\end{eqnarray}
This result is independent of the spectral chirp parameter $b_2$ because the total energy of the laser pulse does not depend on the spectral chirp parameter $b_2$. For the parameters of our numerical example,  (\ref{interpret10}) leads to a value of 0.927, which is in excellent agreement to our numerical results. \newline \newline 
Next, we discuss the norm deviations caused by the oscillatory orders $N_{n,2m,S}^k, k < 2m \leq 2k$. For longer propagation times $t$ these orders make up the dominant contribution to the norm deviations for the simple algorithm because in taking the ratio between the stationary and oscillatory orders (see  (\ref{13Sold}) and (\ref{14Sold})), this number scales with $ (\Delta t/t)^m$ thus for large $t$ the stationary terms are negligible. 

As was discussed in \cite{Renziehausen09JPA}, the oscillatory terms, in leading order, do not depend on the time step $\Delta t$. Therefore we can conclude that the oscillatory orders $N_{n,2m,S}^k, k < 2m \leq 2k$ in contrast to the stationary orders $N_{n,2m,S}^k, k \geq 2m > 0$ do depend on the $\Delta t$-independent part of the difference of the approximative wavefunction $\vec \Psi_S(n,k)$ to the exact wavefunction $\vec \Psi(t)$, which is given in Eqn. (\ref{globalerror2}) by $\vec \phi(t,k)$.   
Futhermore, for the $\Delta t$-independent part of the oscillatory terms on the norm deviations, the following argumentation holds: \newline
The factor $m \left [k! \left (2m-1-k \right)! \right] $ in the denominator of the ratio in  (\ref{14Sold}) increases for an increment of the parameter $m$ of an oscillatory order $N_{n,2m,S}^k$, $k<2m \leq 2k$. However, the term $t^{2m}$ in the nominator of these equations overcompensates this factor for large values of the propagation time $t$. Thus, the larger the order parameter $m$ is, the larger has to be the propagation time $t$, in order that the contribution of the corresponding oscillatory order  $N_{n,2m,S}^k$, $k<2m \leq 2k$ leads of all oscillatory orders to the largest modification of the norm deviations form unity. Due to the factor $(-1)^{k-m}$ in (\ref{14Sold}), this modification of the total norm deviation is negative or positive. Moreover, the sign of the terms alternates as a function of $m$, so that with increasing propagation time $t$, the sign of the contribution of the oscillatory orders to the norm deviations changes. This is the actual reason why we named these terms oscillatory orders in  \cite{Renziehausen09JPA}. \newline
Eventually, for large propagation times $t$, we get due to the positive sign of the highest oscillatory order ($m=k$)
\begin{eqnarray}
N_{t,2k,S}^k = \frac{ t^{2k} }{\left (k! \right)^2} \overline{W}^{2k} +  \mathcal{O} \left ( \Delta t \right )   \label{interpret11} 
\end{eqnarray}
a divergence towards $+\infty$, since the exponent $2m$ in (\ref{14Sold}) reaches its biggest value for $m=k$. Nonetheless, we can retard or even suppress the point in time when the norm deviations from unity caused by the oscillatory orders become a relevant contribution by using a higher value for $k$ in the simulation. This retardation can be seen in figure \ref{fig3alt}, where the use of $k=14$ instead of $k=6$ leads to a later point in time for the divergence. This effect can be interpreted as follows: \newline
The order $k$ represents the maximal number of photons, which interact with the molecule in the numerical simulation. Since the norm deviations caused by the oscillatory terms in leading order do not dependent on the time step $\Delta t$ but on the order $k$ (see (\ref{14Sold})), the norm deviations from unity caused by the oscillatory orders $N_{n,2m,S}^k$, $k<2m \leq 2k$ are related to the truncation of the expansion of the wave function $\vec \Psi_S(n,k)$ for higher orders than the $k^{\textnormal{th}}$-order of the multi-photon processes. Since for the simple algorithm both norm deviations related to the stationary and to the oscillatory orders appear, we have to differentiate in the use of the simple algorithm between these two types of contributions to the norm deviations.

As a next task, we show that from the results presented in Sec. \ref{error_norm} we can conclude that the oscillatory orders $N_{n,2m,S}^k, 2k \geq 2m > k$ (in contrast to the stationary orders) depend on the shape of the potential surfaces $V_j(R)$. In order to explain this, we calculate explicitly with (\ref{orderfunc2}) and (\ref{normexpand5}) the oscillatory orders for the simple algorithm:
\begin{eqnarray}
\hspace{-2,5cm} N_{n,2m, S}^k &=& \sum_{j = 2m-k}^{k} (-1)^{m-j} \Delta t^{2m} \sum_{\mathcal P_{\vec \nu^{(n,j)}}} \sum_{\mathcal P_{\vec \rho^{(n,2m-j)}}}  \left \langle \vec \Psi(0) \left  \vert \prod_{q=1}^{n} \left ( e^{ i \hat H_0 \Delta t} \hat W(q)^{\nu_q^{(n,j)} } \right ) \; \times \right. \right.  \nonumber \\
\hspace{-2,5cm}                 && \left. \left. \times \; \prod_{p=0}^{n-1} \left ( \hat W(n-p)^{\rho_{n-p}^{(n,2m-j)}} e^{ -i \hat H_0 \Delta t } \right) \right \vert \vec \Psi(0) \right \rangle, \label{interpret12}
\end{eqnarray}
The bracket terms $ \left \langle \Psi(0) \left \vert \cdots \right \vert \Psi(0) \right \rangle $  in (\ref{interpret12}) describe the overlap between a bra-state that is influenced by interaction operators $\hat W(q)$ at certain points in time $ \left ( \nu_q^{(n,j)} \neq 0 \right)$ with a ket-state that is influenced by interaction operators $\hat W(n-q)$ at other points in time $ \left ( \rho_{n-q}^{(n,2m-j)} \neq 0 \right)$. Through the impact of the interaction operators at different points in time the bra-state and the ket-state propagate over the time interval $\left [t_0,t_n \right ]$ in a different way. 

That means for points in time $t_{n'}$ with $ t_0 < t_{n'} < t_n $ the bra-state and the ket-state can be localized in different electronic states. But at $t_n$, when the overlap is calculated, the bra-state and the ket-state must be in the same electronic state, otherwise the overlap is zero. If the absolute value of the difference between the potential gradients in the two electronic states  $ \Delta V'(R) := \left \vert \frac {dV_{1}(R)} {dR} - \frac {dV_{0}(R)} {dR} \right \vert $ is large in the spatial region, where the wave function is positioned during the interaction of the molecular system with the laser pulse, the differences in the temporal propagation in these two electronic states $ \vert 1 \rangle$ and $ \vert 0 \rangle$ lead to a small overlap of the bra-state and the ket-state. This leads to the effect that the oscillatory orders are suppressed for large potential gradient differences $\Delta V'(R)$. For this effect we presented already in \cite{Renziehausen09JPA} numerical evidence without analytical proof. In order to comprehend this evidence, we look at figure \ref{fig5alt} again: For an increase of $\Delta V'(R) = 2 m_0$ (see  (\ref{potgrad})) the divergence related to the oscillatory orders is retarded temporally, and for the largest value  $m_0 = 3*10^{-3} a.u.$ the divergence can even be suppressed. 
This effect can interpreted in that way: For a large potential gradient the interaction time for the wave packet in the electronic state $\vert 1\rangle$ with the laser pulse for a transition to the electronic state  $\vert 0 \rangle$ is small and then this interaction can be discribed by a smaller amount of photons. 
As a result, in contrast to the stationary orders not only the electric field $E(t)$ but as well the full dynamics of the molecular system are relevant for the values of the oscillatory orders $N_{n,2m,S}^k, 2k \geq 2m > k$. 

Having discussed the dependence of the oscillatory orders on the potential surfaces $ V_j(R), \; j \in \lbrace 0,1\rbrace$, the dependence on the electric field $E(t)$ leads to interesting phenomena, as well. Here in this context, we mention an effect that we discussed in \cite{Renziehausen09JPA} in detail: \newline If the electric field $E(t)$ is modified under the condition that the total energy of the laser pulse remains unchanged, what leaves the norm deviations caused by the stationary orders unchanged in the limit $ t \rightarrow \infty$ in good approximation, this is not true for the oscillatory orders. Namely, we showed in \cite{Renziehausen09JPA} that the energy conserving temporal broadening of the laser pulse by the introduction of a spectral chirp $b_2$ leads to an enlargement of the norm deviations caused by the oscillatory orders and advantages norm divergences. This means that one needs because of this broadening of the Gaussian laser pulse processes where more photons are exchanged between the laser pulse and the molecular system in order to describe the interaction of the laser pulse with the molecular system in an adequate accuracy. Thus, this example reveals how the results explained here and in  \cite{Renziehausen09JPA} can be used to gain deeper insight about the interaction of a molecular system and a laser pulse. \newline \newline Let us complete the discussion above by the following consideration: Here, our aim is to illustrate why the truncation of the perturbative expansion, taking only $k$ orders in  (\ref{orderfunc}) into account, is a valid ansatz for an approximation, although this expansion decomposes the wave function into terms, which are not orthogonal $ \left ( \left \langle \vec \Psi_{p,S} \vert \vec \Psi_{q,S} \right \rangle \nsim \delta_{pq} \right)$. This illustration is done by a comparison of our situation with a gedankenexperiment, where a wave function is also split into non-orthogonal components. \newline 
Therefore we imagine a transmission grating, that is irradiated with an electron beam, which has a Gaussian density profile. Due to the transmission grating, the wave function of the electrons after the transition through the grating is split into different components: Namely each of the slits creats a circular wave function, and the amplitude of each circular wave is related to the position of the according slit in the Gaussian density profile before the grating. The different circular wave functions interfere, and behind the grating is a screen where the interference pattern caused by the electrons hitting the screen is monitored. Then we arrange an aperture in front of the transmission grating, which is centered at the density profile of the electron beam. Now we adjust the minimal opening of the aperture, so that due to our accuracy of measurement we detect no difference of the interference pattern at the screen between the situation with the aperture and without it. So the result can be interpreted as follows: \newline
The interference pattern is created in good approximation both for the presence and the absence of the aperture by electrons which pass simultaneously through all the slits, which are for the presence of the aperture not masked by it, and we can conclude that because of the fast convergence of the Gaussian profile to zero in good approximation the electrons never pass all the other slits. So it makes no difference for the interference pattern if we mask them by the aperture or not. For the construction of the pattern at the screen, we must not regard the different circular waves of the different slits independently of each other, but we have to take interference effects into account.

In the same sense we can say that if we need $k$ orders to suppress the norm deviations related to the oscillatory orders in our calculations, in the interaction of the laser pulse with the molecular system $0,1,2, \ldots, k$ photon processes happen simultaneously but in good approximation processes which need more than $k$ photons do not happen. 
For such an interpretation of the norm deviations related to the oscillatory orders, we have to take into account that the different perturbation orders of the wave function are not orthogonal and interfere like the circular wave functions of the slits in the scattering experiment; in particular we cannot give an assertion with what probability $\mathcal P(x)$ a specific amount $x$ of photons  with  $  0 \leq x  \leq k$ participates in the laser pulse-molecule interaction.

The norm deviations related to the stationary orders are not relevant for this consideration because they are only norm errors related to the numerical inevitable discretization of time.
\section{\label{sum} Summary} \noindent
In \cite{Renziehausen09JPA} we presented methods for the numerical application of perturbation theory with the aim to characterize the interaction of shaped laser-pulses with molecules. In order to control the quality of the calculations we analysed norm deviations caused by the necessary discretisation of the appearing time-integrals and also the truncation of the perturbation expansion; both causes can lead to substantial deviations. Moreover, in \cite{Renziehausen09JPA} numerical results on a model system incorporating a single nuclear degree of freedom were presented, where a chirped shaped laser pulse induces electronic transitions. These results, in particular the dependence of the different
contributions to the norm deviation on parameters like the propagation
time step, the steepness of the potential curves and the chirp
parameter are explained in \cite{Renziehausen09JPA} for brevity by 
some equations and interrelationships without proofs. The latter are provided in the present paper for what we called simple algorithm in \cite{Renziehausen09JPA}, where the time-integral over one time step occurring in perturbation theory is approximated by one term. \newline
In our analysis of norm deviations from unity calculated with the simple algorithm in \cite{Renziehausen09JPA}, we stated that two classes of terms of different character contribute to the norm deviation: 
The first ones are called the \textit{stationary orders} and are of purely numerical 
nature. They can be suppressed in the limit of small time steps. 
The second kind of contributions, called the \textit{oscillatory orders}, are related to the property of time-dependent perturbation theory, which is not norm conserving. These orders can cause oscillations in the norm of the total wave function, and moreover, for long enough propagation times, these terms can lead to divergences of the norm towards infinity. In this publication, we proof equations and interrelations which explain this behavior of the stationary and the oscillatory orders for the simple algorithm that was documented for numerical calculated results in \cite{Renziehausen09JPA}. Moreover, here we present a method to calculate the norm deviations caused by the stationary orders quantitatively analytically. The accuracy of this method is demonstrated for a numerical example.
 
In \cite{Renziehausen09JPA} we stated furthermore that the oscillatory terms directly correlate with the order of the multi-photon transitions to be described, and by increasing the order of the perturbation theory the
norm deviations can be reduced. In this publication we generate, beyond that, an easy interpretable graphic image of this reduction. Therefore, we compare our situation to the situation that a transmission grating is irradiated by an electron beam with a Gaussian density profile, what causes an interference pattern on a screen behind the grating. Moreover, there is in front of the grating an aperture centered around the electron beam, that we open further until the interference pattern does not change anymore.

In the future, we will publish in \cite{Renziehausen11JMP} how the calculation done in this paper can be devolved from the simple algorithm to what in \cite{Renziehausen09JPA} we called improved algorithm. Moreover, our research goals are to implement the results depicted in \cite{Renziehausen09JPA} and in this publication in order to get a better understanding of multi-photon processes. With the help of the derived results we are now in the position to analyse which processes of different orders are relevant for the correct description of a chemical reaction. Therefore we will compare the results calculated with converged perturbation theory and numerically exact results, which contain all perturbation orders.

\section*{Acknowledgment} \noindent
Financial support by the Freistaat Bayern within the BayEFG and by the DFG within the Graduiertenkolleg 1221 is gratefully acknowledged. Many stimulating discussions with Volker Engel, Martin Brüning, Sarah Caroll Galleguillos Kempf and Gunther Dirr are acknowledged. 

\begin{appendix}
\section{\label{appendix A} Proof of  (\ref{wavefunctionS}) for the wave functions $\vec \Psi_S(n,k)$}
In this appendix we prove that the wave functions $\vec \Psi_{S}(n,k)$ for the simple algorithm can be written in the form of  (\ref{wavefunctionS}), which is
\begin{eqnarray}
\hspace{-1.5 cm} \vec \Psi_{S}(n,k) = \left[ \sum_{m=0}^k (-i\; \Delta t)^m \sum_{\mathcal P_ {\vec \nu^{(n,m)}}} \prod_{j=0}^{n-1} \left( \hat W(n-j)^{\nu_{n-j}^{(n,m)}} e^{-i \hat H_0 \Delta t} \right) \right] \vec \Psi(0), \nonumber
\end{eqnarray}
where we use the definition (\ref{twelve}) for the simple algorithm as a starting point for the proof. \newline
\newline 
As an inception of the proof we calculate as helpful lemmas with  (\ref{twelve}) the following equations for $\vec \Psi_S(1,k)$ and $\vec \Psi_S(n,0)$:
\begin{eqnarray}
\hspace{-1.5 cm} \vec \Psi_S(1,k) &=& e^{-i \hat H_0 \Delta t}\vec \Psi(0)-i \; \Delta t \; \hat W(1)\vec \Psi_S(1,k-1) \nonumber \\
\hspace{-1.5 cm}                &=& \sum_{m=0}^k (-i \; \Delta t)^m \hat W(1)^m e^{-i \hat H_0 \Delta t} \vec \Psi(0) \nonumber \\
\hspace{-1.5 cm}                 &=& \left[ \sum_{m=0}^k (-i\; \Delta t)^m \sum_{\mathcal P_ {\vec \nu^{(1,m)}}} \prod_{j=0}^{0} \left( \hat W(1-j)^{\nu_{1-j}^{(1,m)}} e^{-i \hat H_0 \Delta t} \right) \right] \vec \Psi(0) \label{lemma1}
\end{eqnarray}
and 
\begin{eqnarray}
\hspace{-1.5 cm} \vec \Psi_S(n,0) &=& e^{-i n \hat H_0 \Delta t} \vec \Psi(0) \nonumber \\
                 &=& \hspace{-0.2cm} \left[ \sum_{m=0}^0 (-i\; \Delta t)^m \sum_{\mathcal P_ {\vec \nu^{(n,m)}}} \prod_{j=0}^{n-1} \left( \hat W(n-j)^{\nu_{n-j}^{(n,m)}} e^{-i \hat H_0 \Delta t} \right) \right] \vec \Psi(0). \label{lemma2} 
\end{eqnarray}
For the proof of (\ref{wavefunctionS}), we work with complete induction and the induction hypothesis
\begin{eqnarray}
\hspace{-2,5cm} \vec \Psi_{S}(p+1 , q-p) &=& \left[ \sum_{m=0}^{q-p} (-i\; \Delta t)^m  \sum_{\mathcal P_ {\vec \nu^{(p+1,m)}}} \prod_{j=0}^{p} \left( \hat W(p+1-j)^{\nu_{p+1-j}^{(p+1,m)}} e^{-i \hat H_0 \Delta t} \right) \right] \vec \Psi(0) \nonumber \\ 
&& \textnormal{for a} \; q  \in \mathbb{N} \; \textnormal{and} \;  p = 0, 1,2,...,q-1.  \label{indassump1} 
\end{eqnarray}
For $p=0$ and an arbitrary $q \in \mathbb{N}$ is the induction hypothesis (\ref{indassump1}) true because of  (\ref{lemma1}). This implies that (\ref{indassump1}) is true for the base case $q=1$. 

The induction hypothesis (\ref{indassump1}) implies by substituting $p$ by $p+1$  the equation 
\begin{eqnarray} 
\hspace{-2,5cm}  \vec \Psi_{S}(p+2,q-p-1) &=& \hspace{-0.2 cm} \left[ \sum_{m=0}^{q-p-1} (-i\; \Delta t)^m \hspace{-0.2 cm} \sum_{\mathcal P_ {\vec \nu^{(p+2,m)}}} \prod_{j=0}^{p+1} \left( \hat W(p+2-j)^{\nu_{p+2-j}^{(p+2,m)}} e^{-i \hat H_0 \Delta t} \right) \right] \vec \Psi(0)  \nonumber \\ 
\hspace{-2,5cm}  && \textnormal{for a} \; q  \in \mathbb{N} \; \textnormal{and} \;  p = 0, 1,2,...,q-2.  \label{zwischenergebnis} 
\end{eqnarray} 
Because of  (\ref{lemma2}) we can assume that (\ref{zwischenergebnis}) is correct for $p=q-1$, too. With the definition (\ref{twelve}) for the simple algorithm we get
\begin{eqnarray} 
\hspace{-2,5cm}  \vec \Psi_S(p+2, q-p) &=& e^{-i \hat H_0 \Delta t} \vec \Psi(p+1, q-p) - i \; \Delta t \; \hat W(p+2) \vec \Psi_S(p+2, q-p-1) \nonumber \\
&& \textnormal{for a} \; q  \in \mathbb{N} \; \textnormal{and} \;  p = 0, 1,2,...,q-1.  \label{zwischenergebnis2}
\end{eqnarray} 
Inserting (\ref{indassump1}) and (\ref{zwischenergebnis}) into (\ref{zwischenergebnis2}) leads to 
\begin{eqnarray}
&&\hspace{-2,5cm}  \vec \Psi_{S}(p+2 , q-p) = \left[ \sum_{m=0}^{q-p} (-i\; \Delta t)^m e^{-i \Delta t \hat H_0} \hspace{-0.4 cm}
\sum_{\mathcal P_ {\vec \nu^{(p+1,m)}}} \prod_{j=0}^{p} \left( \hat W(p+1-j)^{\nu_{p+1-j}^{(p+1,m)}} e^{-i \hat H_0 \Delta t} \right) \right. \nonumber \\
&& \hspace{-2,3cm} \left. +  \hspace{-0.1 cm} \sum_{m=0}^{q-p-1} (-i\; \Delta t)^{m+1} \hat W(p+2) \hspace{-0.4 cm} \sum_{\mathcal P_ {\vec \nu^{(p+2,m)}}} \prod_{j=0}^{p+1} \left( \hat W(p+2-j)^{\nu_{p+2-j}^{(p+2,m)}} e^{-i \hat H_0 \Delta t} \right) \right]  \hspace{-0.05 cm}  \vec \Psi(0)  \nonumber
 \nonumber \\ 
&&\hspace{-2,6cm} = \; \left[ \sum_{m=0}^{q-p} (-i\; \Delta t)^m e^{-i \Delta t \hat H_0} \hspace{-0.4 cm} \sum_{\mathcal P_ {\vec \nu^{(p+1,m)}}} \prod_{j=0}^{p} \left( \hat W(p+1-j)^{\nu_{p+1-j}^{(p+1,m)}} e^{-i \hat H_0 \Delta t} \right) \right. \nonumber \\
&& \hspace{-2,3cm} \left. +  \sum_{m=1}^{q-p} (-i\; \Delta t)^{m} \hat W(p+2) \hspace{-0.4 cm} \sum_{\mathcal P_ {\vec \nu^{(p+2,m-1)}}} \prod_{j=0}^{p+1} \left( \hat W(p+2-j)^{\nu_{p+2-j}^{(p+2,m-1)}} e^{-i \hat H_0 \Delta t} \right) \right] \hspace{-0.05cm} \vec \Psi(0)  \label{zwischenergebnis3} \\
&\hspace{-2,6cm} =& \hspace{-2,0cm} \left[ \sum_{m=0}^{q-p} (-i\; \Delta t)^m \hspace{-0.4 cm}  \sum_{\mathcal P_ {\vec \nu^{(p+2,m)}}} \prod_{j=0}^{p+1} \left( \hat W(p+2-j)^{\nu_{p+2-j}^{(p+2,m)}} e^{-i \hat H_0 \Delta t} \right) \right] \hspace{-0.1cm} \vec \Psi(0) \label{zwischenergebnis4}  \\
&& \hspace{-2,0cm} \textnormal{for a} \; q  \in \mathbb{N} \; \textnormal{and} \;  p = 0, 1,2,...,q-1.  \nonumber
\end{eqnarray}
In this calculation we regarded the fact that in (\ref{zwischenergebnis3}) the term in the upper line contains only terms without the operator $\hat W(p+2)$ and the term in the lower line contains only terms where $\hat W(p+2)$ appears at least in first power in order to see that the terms in (\ref{zwischenergebnis3}) and (\ref{zwischenergebnis4}) are identical. Now we substitute in (\ref{zwischenergebnis4}) $p$ by $p-1$ and get:
\begin{eqnarray}
\hspace{-2,5cm}  \vec \Psi_S(p+1,q+1-p) &=& \hspace{-0.2 cm} \left[ \sum_{m=0}^{q+1-p} (-i\; \Delta t)^m \hspace{-0.2cm} \sum_{\mathcal P_ {\vec \nu^{(p+1,m)}}} \prod_{j=0}^{p} \left( \hat W(p+1-j)^{\nu_{p+1-j}^{(p+1,m)}} e^{-i \hat H_0 \Delta t} \right) \right] \vec \Psi(0) \nonumber \\
 && \textnormal{for a} \; q  \in \mathbb{N}  \; \textnormal{and} \;  p = 1,2,...,q  \label{zwischenergebnis5}
\end{eqnarray}
Moreover (\ref{lemma1}) implies that (\ref{zwischenergebnis5}) is true for $p=0$, too and this means that we have made the induction step and (\ref{indassump1}) is true for all $ q  \in \mathbb{N}$. With the substitutions $n=p+1$ and $k=q-p$ in  (\ref{indassump1}) the formula for the wave functions  $\vec \Psi_{S}(n,k)$ for the simple algorithm  (\ref{wavefunctionS}) is proven.  $\square$

\section{\label{appendix C} Proof of  (\ref{statSRule}) for the stationary orders of the simple algorithm}
In this appendix we prove that for the stationary orders of the simple algorithm (\ref{statSRule}) holds:
\begin{eqnarray}
N_{n,2m, S}^k = (-1)^m \Delta t^{2m} \sum_{\mathcal P_{\vec \nu^{(n,m)}}} \prod_{j=1}^n W(j)^{2 \; \nu_j^{(n,m)}} \nonumber 
\end{eqnarray}
For the proof of (\ref{statSRule}) we work with complete induction over $n$ for an arbitrary $m$ that fulfils the condition  $k\geq 2m >0$. \newline \newline 
As a proof for the base case $n=1$, we derive with  (\ref{normexpand6}) that for the norm orders $N_{1,2m, S}^k$  (\ref{statSRule}) holds:
\begin{eqnarray}
\hspace{-2,5cm}  N_{1,2m, S}^k &=& \sum_{j = 0}^{2m} \left \langle \vec \Psi_{j,S}(1)\vert \vec \Psi_{2m-j,S}(1) \right \rangle \nonumber \\ 
\hspace{-2,5cm}              && \hspace{-1,5cm} = \; \sum_{j = 0}^{2m} \left \langle \vec \Psi(0) \vert e^{i \hat H_0 \Delta t} \hat W(1)^j \Delta t^j i^j (-i)^{2m-j} \Delta t^{2m - j} \hat W(1)^{2m-j} e^{-i \hat H_0 \Delta t} \vert \vec \Psi(0) \right \rangle  \nonumber \\
\hspace{-2,5cm}              && \hspace{-1,5cm} = \; (-1)^m \Delta t^{2m} W(1)^{2m} \nonumber \\
\hspace{-2,5cm}              && \hspace{-1,5cm} = \;  (-1)^m \Delta t^{2m} \sum_{\mathcal P_{\vec \nu^{(1,m)}}} \prod_{j=1}^1 W(1)^{2\nu_j^{(1,m)}} \surd 
\end{eqnarray}
Then as an induction hypothesis we suppose that all norm orders $N_{\eta,2m,S}^k$ for $ n \leq \eta \leq 1$ with an arbitrarily  chosen $n \in \mathbb{N}$ suffice  (\ref{statSRule}) and show as induction step that this implies that $N_{n+1,2m,S}^k$ suffices  (\ref{statSRule}), too. For this purpose we derive with  (\ref{normexpand6}) first that for $N_{n+1,2m,S}^k$ holds 
\begin{eqnarray}
\hspace{-2,5cm}  N_{n+1,2m, S}^k &=& \sum_{j = 0}^{2m} \left \langle \vec \Psi_{j,S}(n+1) \vert \vec \Psi_{2m-j,S}(n+1) \right \rangle \nonumber \\ 
 \hspace{-2,5cm}               && \hspace{-1,5cm}  = \; \sum_{j = 0}^{2m} \Delta t^{2m} \sum_{\mathcal P_{\vec \nu^{(n+1,j)}}} \sum_{\mathcal P_{\vec \rho^{(n+1,2m-j)}}}  \left \langle \vec \Psi(0) \left  \vert i^j \prod_{q=1}^{n+1} \left ( e^{ i \hat H_0 \Delta t} \hat W(q)^{\nu_q^{(n+1,j)} } \right ) \right. \right. \times \nonumber \\
\hspace{-2,5cm}                && \hspace{-1,0cm} \left. \left.  \times \;(-i)^{2m-j} \prod_{p=0}^n \left ( \hat W(n+1-p)^{\rho_{n+1-p}^{(n+1,2m-j)}} e^{ -i \hat H_0 \Delta t } \right) \right \vert \vec \Psi(0) \right \rangle \label{prostatS1}
\end{eqnarray}
Now we write out the operator $\hat W(n+1)$ explicitly 
\begin{eqnarray}
\hspace{-2,5cm}  N_{n+1,2m, S}^k &=&  (-1)^{m} \Delta t^{2m} \sum_{j = 0}^{2m} \sum_{g=0}^j \sum_{f=0}^{2m-j} \sum_{\mathcal P_{\vec \nu^{(n,j-g)}}} \sum_{\mathcal P_{\vec \rho^{(n,2m-j-f)}}}  (-1)^j \; \times \nonumber \\
 \hspace{-2,5cm}   &&  \hspace{-1,0cm}  \times \; \left \langle \vec \Psi(0) \left  \vert \prod_{q=1}^{n} \left ( e^{i \hat H_0 \Delta t} \hat W(q)^{\nu_q^{(n,j-g)}} \right )  e^{i \hat H_0 \Delta t} \; \hat W(n+1)^{g+f} \;  e^{-i \hat H_0 \Delta t} \right. \right. \times \nonumber \\
 \hspace{-2,5cm}                  &&  \hspace{-1,0cm}  \hspace{-0,1cm} \left. \left. \times \;\prod_{p=0}^{n-1} \left ( \hat W(n-p)^{\rho_{n-p}^{(n,2m-j-f)}} e^{-i \hat H_0 \Delta t } \right) \right \vert \vec \Psi(0) \right \rangle, \label{prostatS2}
\end{eqnarray}
and then we introduce the new sum indices $s:=g+f$ and $d:=j-g$:
\begin{eqnarray}
\hspace{-2,5cm}  N_{n+1,2m, S}^k &=& (-1)^m \Delta t^{2m} \sum_{j = 0}^{2m} \sum_{d=0}^j \sum_{s=j-d}^{2m-d} \sum_{\mathcal P_{\vec \nu^{(n,d)}}} \sum_{\mathcal P_{\vec \rho^{(n,2m-d-s)}}} (-1)^j \; \times \nonumber \\
\hspace{-2,5cm} && \hspace{-1,0cm} \times \; \left \langle \vec \Psi(0) \left \vert \prod_{q=1}^{n} \left ( e^{i \hat H_0 \Delta t} \hat W(q)^{\nu_q^{(n,d)}} \right ) e^{i \hat H_0 \Delta t} \; \hat W(n+1)^{s} \;  e^{-i \hat H_0 \Delta t} \; \times \right. \right. \nonumber \\
\hspace{-2,5cm}   && \hspace{-1,0cm}  \hspace{-0,1cm} \left. \left. \times \; \prod_{p=0}^{n-1}  \left ( \hat W(n-p)^{ \rho_{n-p}^{(n,2m-d-s)} } e^{-i \hat H_0 \Delta t } \right) \right \vert \vec \Psi(0) \right \rangle \label{prostatS3}
\end{eqnarray} 
\setlength{\unitlength}{1.0cm} \begin{figure}   [b!]
\centering  
\resizebox{0.6 \textwidth}{!}{\includegraphics[width=10.5cm,height=10.5cm]{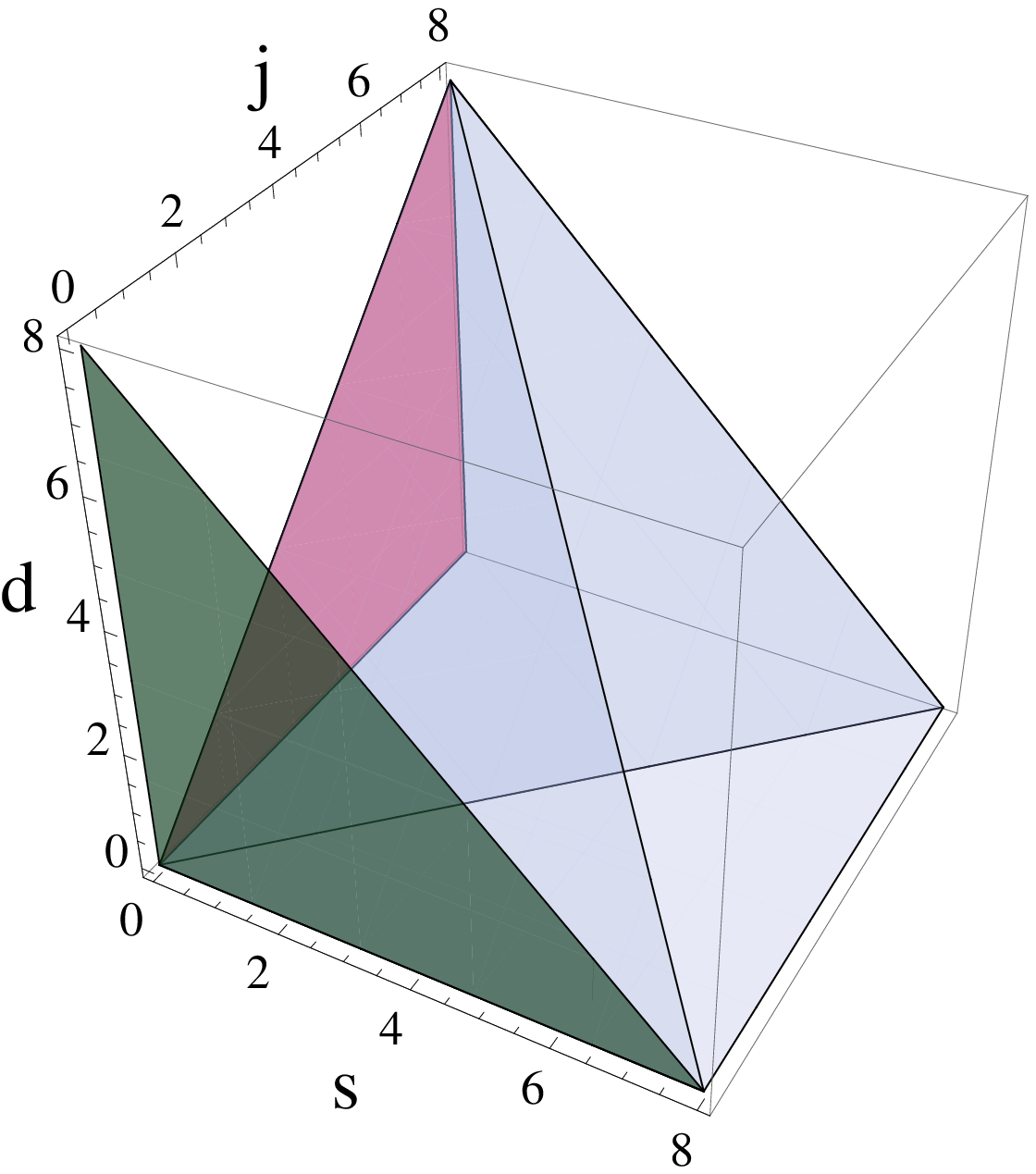}}
\caption{
The pyramid, that we use for the visualization of the threefold sums $\sum_{j = 0}^{2m} \sum_{d=0}^j \sum_{s=j-d}^{2m-d}$ on the left side and  accordingly  $\sum_{s = 0}^{2m}  \sum_{d=0}^{2m-s} \sum_{j=d}^{d+s} $ on the right side of (\ref{prostatS3.1}), is presented here for $2m = 8$. Visualization means as explained in the text that all 3-tuples $(j,d,s)$ that are inside or on the boundary of this pyramid, which is given by the four points $(0,0,0), \; (0,0,8), \; (8,0,8), \; (8,8,0)$, represent exactly all the combinations for $j, d$ and $s$ that are contained in these both threefold sums $\sum_{j = 0}^{8} \sum_{d=0}^j \sum_{s=j-d}^{8-d} $ and  $\sum_{s = 0}^{8}  \sum_{d=0}^{8-s} \sum_{j=d}^{d+s}$. Moreover, we have drawn in this figure the projections of this pyramid on the $(j,d)$-plane and the $(s,d)$-plane, which are triangles.} \label{fig1} \end{figure}
\hspace{-0.45 cm} Now we think about (\ref{prostatS3}) and cognize that there appears the threefold sum $\sum_{j = 0}^{2m} \sum_{d=0}^j \sum_{s=j-d}^{2m-d}$. For a threefold sum of this form and an arbitrarily function $\zeta(j,d,s)$ the equation 
\begin{eqnarray}
\sum_{j = 0}^{2m} \sum_{d=0}^j \sum_{s=j-d}^{2m-d} \zeta(j,d,s) = \sum_{s = 0}^{2m}  \sum_{d=0}^{2m-s} \sum_{j=d}^{d+s} \zeta(j,d,s) \label{prostatS3.1}
\end{eqnarray}
holds. The Eqn. (\ref{prostatS3.1}) can be visualized by the picture that all 3-tuples $(j,d,s)$ that are inside or on the boundary of a pyramid which is given by the four points $(0,0,0), \; (0,0,2m), \; (2m,0,2m), \; (2m,2m,0)$, represent exactly all the combinations for $j, d$ and $s$ that are contained both in the threefold sum $\sum_{j = 0}^{2m} \sum_{d=0}^j \sum_{s=j-d}^{2m-d}$ on the left side of  (\ref{prostatS3.1}) and in the threefold sum $\sum_{s = 0}^{2m}  \sum_{d=0}^{2m-s} \sum_{j=d}^{d+s}$ on the right side of (\ref{prostatS3.1}), thus, these sums are equal. The pyramid is presented in figure \ref{fig1} for $2m=8$. The visualization can explained in detail as follows: \newline

\begin{figure}  
\centering  
\resizebox{0.6 \textwidth}{!}{\includegraphics[width=10.5cm,height=10.5cm]{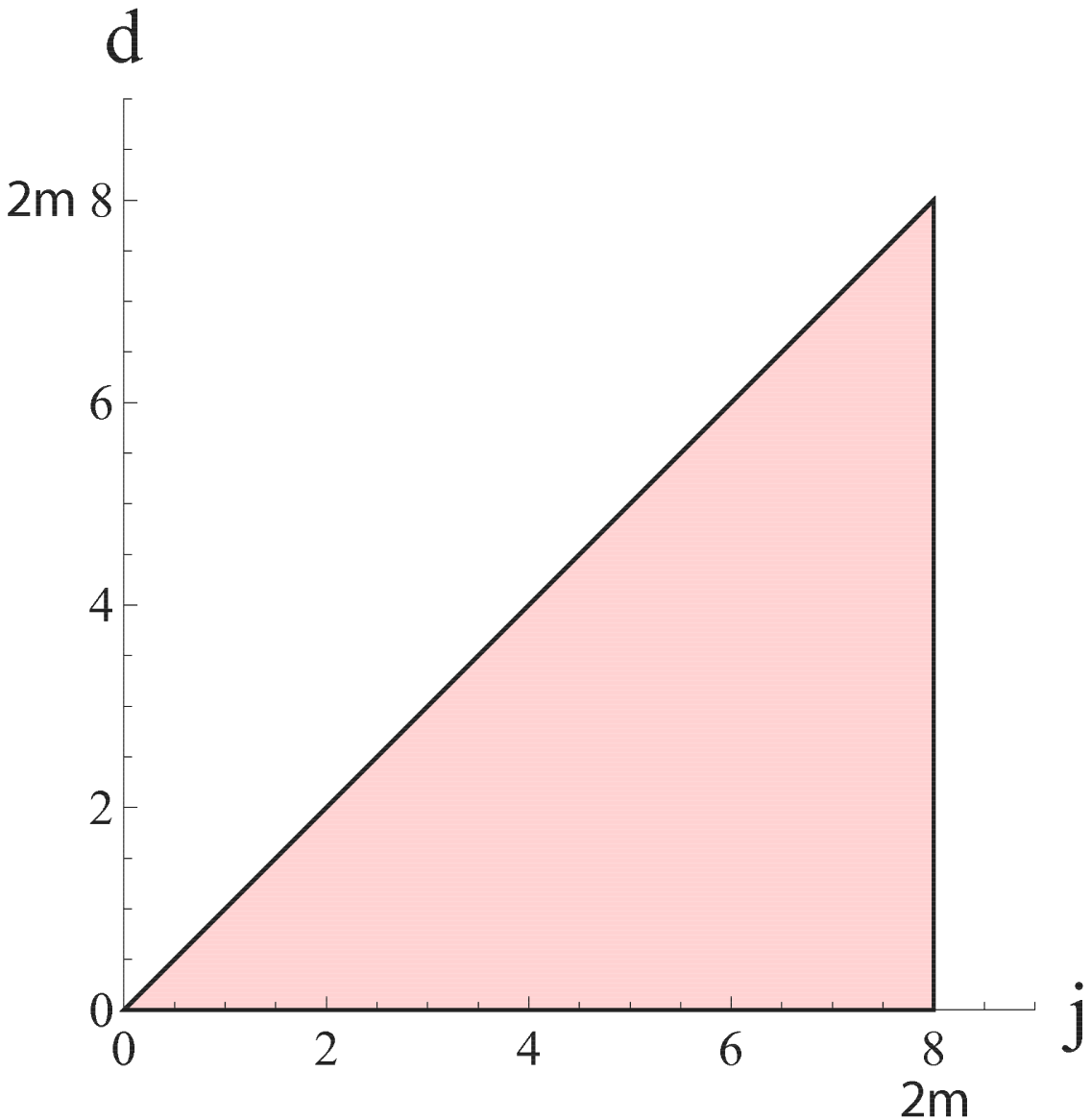}}
\caption{The triangle, that we use for the visualization of the outer twofold sum $\sum_{j = 0}^{2m} \sum_{d=0}^j$ over $j$ and $d$ on the left side of (\ref{prostatS3.1}), is presented here for $2m = 8$. We obtain this triangle by a projection of the pyramid in figure \ref{fig1} on the $(j,d)$-plane.} 
\label{fig2}
\end{figure}

For the visualization of the threefold sum on the left side of (\ref{prostatS3.1}) it is helpful to think first about the projection of the pyramid on the $j-d-$plane; this is a triangle, which is given in figure \ref{fig2} for $2m=8$:
\setlength{\unitlength}{1.0cm} 

All 2-tuples $(j,d)$ that are inside or on the boundary of this triangle represent exactly all the combinations for $j$ and $d$ that are contained in the twofold sum $\sum_{j = 0}^{2m} \sum_{d=0}^j$. When we combine this result with the question which values for $s$ for a given 2-tuple $(j,d)$ inside or on the boundary of the triangle shown in figure \ref{fig2} induce 3-tuples $(j,d,s)$ that correspond to points located in or on the boundary of the pyramid in figure \ref{fig1}, we can comprehend that all 3-tuples $(j,d,s)$, that are inside or on the boundary of this pyramid represent exactly all the combinations for $j, d$ and $s$ that are contained in the threefold sum $\sum_{j = 0}^{2m} \sum_{d=0}^j \sum_{s=j-d}^{2m-d}$ on the left side of  (\ref{prostatS3.1}). 

\begin{figure}  
\centering 
\resizebox{0.6 \textwidth}{!}{\includegraphics[width=10.5cm,height=10.5cm]{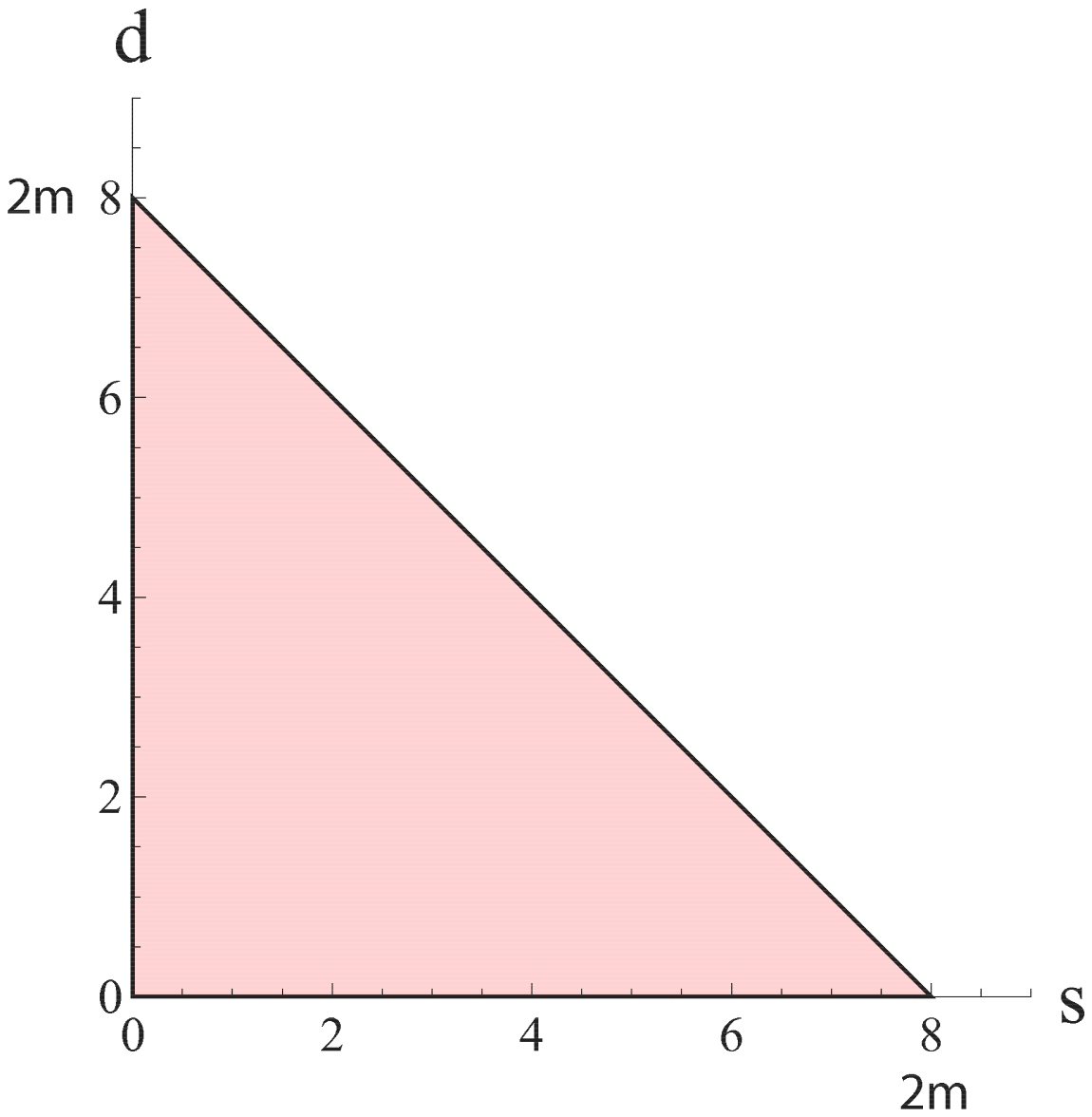}}
\caption{The triangle, that we use for the visualization of the outer twofold sum $\sum_{s = 0}^{2m} \sum_{d=0}^{2m-s}$ over $s$ and $d$ on the right side of (\ref{prostatS3.1}), is presented here for $2m = 8$. We obtain this triangle by a projection of the pyramid in figure \ref{fig1} on the $(s,d)$-plane.} 
\label{fig3}
\end{figure}

By an analogue approach for the visualization of the threefold sum on the right side of (\ref{prostatS3.1}) it is helpful to think first about the projection of the pyramid on the $s-d-$plane; this is a triangle, which is given in figure \ref{fig3} for $2m =8$:
\setlength{\unitlength}{1.0cm} 
All 2-tuples $(s,d)$ that are inside or on the boundary of this triangle represent exactly all the combinations for $s$ and $d$ that are contained in the twofold sum $\sum_{s = 0}^{2m} \sum_{d=0}^{2m-s}$. When we combine this result with the question, which values for $j$ for a given 2-tuple $(s,d)$ inside or on the boundary of the triangle shown in figure \ref{fig3}, induce 3-tuples $(j,d,s)$ that correspond to points located in or on the boundary of the pyramid in figure \ref{fig1}, we can comprehend that all 3-tuples $(j,d,s)$, that are inside or on the boundary of this pyramid represent exactly all the combinations for $j, d$ and $s$ that are contained in the threefold sum $\sum_{s = 0}^{2m} \sum_{d=0}^{2m-s} \sum_{j=d}^{d+s}$ on the right side of  (\ref{prostatS3.1}). \newline
Now for (\ref{prostatS3}) the summands  $\zeta(j,d,s)$ in the threefold sum over $j, d$ and $s$ are of the special form 
\begin{eqnarray}
 \zeta(j,d,s) = (-1)^j \phi (d,s) \label{prostatS3.2}
\end{eqnarray}
and moreover, it is to show straightforwardly that 
\begin{eqnarray}
\sum_{j=p}^{q} (-1)^j &=& (-1)^p \delta_{(q-p) \; mod \; 2, 0}, \label{prostatS3.3} \\
\sum_{j=p}^{q} (-1)^j &=& (-1)^q \delta_{(q-p) \; mod \; 2, 0}. \label{prostatS3.4}
\end{eqnarray}
In the following calculations in this paper we will use both (\ref{prostatS3.3}) and  (\ref{prostatS3.4}). Thus, for a function  $\zeta(j,d,s)$ for that (\ref{prostatS3.2}) is valid, one can conclude with  (\ref{prostatS3.1}) and (\ref{prostatS3.3}) that holds
\begin{eqnarray}
\sum_{j = 0}^{2m} \sum_{d=0}^j \sum_{s=j-d}^{2m-d} (-1)^j \phi (d,s)  &=& \sum_{s = 0}^{2m}  \sum_{d=0}^{2m-s} \sum_{j=d}^{d+s}  (-1)^j \phi (d,s) \nonumber \\
&=& \sum_{s = 0}^{2m}  \sum_{d=0}^{2m-s} (-1)^d \; \delta_{s \; mod \; 2,0} \; \phi (d,s) \nonumber \\
&=& \sum_{r = 0}^{m} \sum_{d=0}^{2m-2r} (-1)^d \phi (d,2r). \label{prostatS3.5}
\end{eqnarray}
With (\ref{prostatS3.5}) we can simplify (\ref{prostatS3}) and get as an intermediate result:
\begin{eqnarray}
\hspace{-2,5cm}  N_{n+1,2m, S}^k &=&  (-1)^m \Delta t^{2m} \sum_{r = 0}^{m} \sum_{d=0}^{2m-2r} \sum_{\mathcal P_{\vec \nu^{(n,d)}}} \sum_{\mathcal P_{\vec \rho^{(n,2m-d-2r)}}} (-1)^d \; \times \nonumber \\
\hspace{-2,5cm} && \times \; \left \langle \vec \Psi(0) \left \vert \prod_{q=1}^{n} \left ( e^{i \hat H_0 \Delta t} \hat W(q)^{\nu_q^{(n,d)}} \right ) e^{i \hat H_0 \Delta t} \; W(n+1)^{2r} \right. \right. \times \nonumber \\
\hspace{-2,5cm}                 && \left. \left. \hspace{-0.06 cm} \times \;   e^{-i \hat H_0 \Delta t} \prod_{p=0}^{n-1}  \left ( \hat W(n-p)^{ \rho_{n-p}^{(n,2m-d-2r)} } e^{- i \hat H_0 \Delta t } \right) \right \vert \vec \Psi(0) \right \rangle \label{prostatS4}
\end{eqnarray} 
Reminding that $W(n+1)^{2r}$ commutates with all operators, we simplify (\ref{prostatS4}):
\begin{eqnarray}
\hspace{-2,5cm}  N_{n+1,2m, S}^k &=& \sum_{r = 0}^{m} (-1)^r \Delta t^{2r} W(n+1)^{2r} \; \times \nonumber \\ 
\hspace{-2,5cm} && \times \; \left [ (-1)^{m-r} \Delta t^{2(m-r)} \sum_{d=0}^{2(m-r)} \sum_{\mathcal P_{\vec \nu^{(n,d)}}} \sum_{\mathcal P_{\vec \rho^{(n,2(m-r)-d)}}} (-1)^{d} \; \times \right. \nonumber \\ 
\hspace{-2,5cm} && \times \; \left. \left \langle \vec \Psi(0) \left \vert \prod_{q=1}^{n} \left ( e^{i \hat H_0 \Delta t} \hat W(q)^{\nu_q^{(n,d)}} \right ) \; \times \right. \right. \right. \nonumber \\
\hspace{-2,5cm} && \left. \left. \left. \hspace{-0.11 cm} \times \; \prod_{p=0}^{n-1}  \left ( \hat W(n-p)^{ \rho_{n-p}^{(n,2(m-r)-d)} }  e^{- i \hat H_0 \Delta t } \right) \right \vert \vec \Psi(0) \right \rangle \right] \nonumber  \\
\hspace{-2,5cm} &=& \sum_{r = 0}^{m} (-1)^r \Delta t^{2r} W(n+1)^{2r} N_{n,2(m-r),S}^k  \label{prostatS5}
\end{eqnarray}
Employing the induction hypothesis, we can substitute now for the terms $N_{n,2(m-r),S}^k$ in (\ref{prostatS5}) the right side of  (\ref{statSRule}) and implicate that (\ref{statSRule}) is true for $n+1$: 
\begin{eqnarray}
\hspace{-1,5cm}  N_{n+1,2m, S}^k &=& \sum_{r = 0}^{m} (-1)^r \Delta t^{2r} W(n+1)^{2r} (-1)^{m-r} \Delta t^{2(m-r)}  \times \; \nonumber \\ \hspace{-2,5cm} && \times
 \sum_{\mathcal P_{\vec \nu^{(n,m-r)}}} \prod_{j=1}^n W(j)^{2 \; \nu_j^{(n,m-r)}} \nonumber \\
\hspace{-1,5cm}                 &=&  (-1)^m \Delta t^{2m} \sum_{\mathcal P_{\vec \nu^{(n+1,m)}}} \prod_{j=1}^{n+1} W(j)^{2 \; \nu_j^{(n+1,m)}} \label{prostatS6}
\end{eqnarray}
Thus, the complete induction proof is succeeded and  (\ref{statSRule}) is proven. $\square$
\section{\label{appendix D} Proof of the annihilation hypothesis for the oscillatory orders of the simple algorithm}
In this appendix we prove the annihilation hypothesis for the oscillatory orders of the simple algorithm, $N_{n,2m,S}^k, k< 2m \leq 2k$. \newline \newline
As a starting point for this proof we take (\ref{prostatS6.1}) with the substitution 
$n \rightarrow n+1$ 
\begin{eqnarray}
&& \hspace{-2,5cm} N_{n+1,2m, S}^k = \Delta t^{2m}  \sum_{j = 2m-k}^{k} \sum_{\mathcal P_{\vec \nu^{(n+1,j)}}} \sum_{\mathcal P_{\vec \rho^{(n+1,2m-j)}}} (-1)^{m-j}  \; \times \nonumber \\
&& \hspace{-1,5cm}\times \;  \left \langle \vec \Psi(0) \left  \vert \prod_{q=1}^{n+1} \left ( e^{ i \hat H_0 \Delta t} \hat W(q)^{\nu_q^{(n+1,j)} } \right ) \prod_{p=0}^n \left ( \hat W(n+1-p)^{\rho_{n+1-p}^{(n+1,2m-j)}} e^{ -i \hat H_0 \Delta t } \right) \right \vert \vec \Psi(0) \right \rangle \nonumber
\end{eqnarray}
and transform it in the same manner we transformed in \ref{appendix C} (\ref{prostatS1}) into (\ref{prostatS3}). 
Thus, we get as a result:
\begin{eqnarray}
&& \hspace{-2,5cm} N_{n+1,2m, S}^k = (-1)^m \Delta t^{2m} \sum_{j = 2m-k}^{k} \sum_{d=0}^j \sum_{s=j-d}^{2m-d} \sum_{\mathcal P_{\vec \nu^{(n,d)}}} \sum_{\mathcal P_{\vec \rho^{(n,2m-d-s)}}} (-1)^j \; \times \nonumber \\ 
&& \hspace{-1,5cm} \times \; \left \langle \vec \Psi(0) \left \vert \prod_{q=1}^{n} \left ( e^{i \hat H_0 \Delta t} \hat W(q)^{\nu_q^{(n,d)}} \right )  e^{i \hat H_0 \Delta t} \; \times \right. \right. \nonumber \\
&& \hspace{-1,5cm} \times \; \left. \left.  \hat W(n+1)^{s} \;  e^{-i \hat H_0 \Delta t}  \prod_{p=0}^{n-1}  \left ( \hat W(n-p)^{ \rho_{n-p}^{(n,2m-d-s)} } e^{-i \hat H_0 \Delta t } \right) \right \vert \vec \Psi(0) \right \rangle \label{prostatS6.2}
\end{eqnarray}
Now we think about  (\ref{prostatS6.2}) and cognize that there appears the threefold sum $\sum_{j = 2m-k}^{k} \sum_{d=0}^j \sum_{s=j-d}^{2m-d}$. For a threefold sum of this form and an arbitrary function $\zeta(j,d,s)$ the equation 
\begin{eqnarray}
\hspace{-2,5cm} \sum_{j = 2m-k}^{k} \sum_{d=0}^j \sum_{s=j-d}^{2m-d} \zeta(j,d,s) = \sum_{s = 0}^{2m}  \sum_{d=max(2m-k-s, \; 0)}^{min(2m-s, \; k)} \sum_{j = max(2m-k, \; d)}^{ min(d+s, \; k)} \zeta(j,d,s) \label{prostatS6.3}
\end{eqnarray}
holds, thus we can transform (\ref{prostatS6.2}) into 
\begin{eqnarray}
&& \hspace{-2,5cm}  N_{n+1,2m, S}^k = (-1)^m \Delta t^{2m} \sum_{s = 0}^{2m}  \sum_{d=max(2m-k-s, \; 0)}^{min(2m-s, \; k)} \sum_{j = max(2m-k, \; d)}^{ min(d+s, \; k)} (-1)^j \; \times \nonumber \\
&& \hspace{-1,5cm}   \times \; \sum_{\mathcal P_{\vec \nu^{(n,d)}}} \sum_{\mathcal P_{\vec \rho^{(n,2m-d-s)}}}  \left \langle \vec \Psi(0) \left \vert \prod_{q=1}^{n} \left ( e^{i \hat H_0 \Delta t} \hat W(q)^{\nu_q^{(n,d)}} \right ) e^{i \hat H_0 \Delta t} \; \times \right. \right. \nonumber \\
&& \hspace{-1,5cm}   \times \; \left. \left.  \hat W(n+1)^{s} \; e^{-i \hat H_0 \Delta t} \prod_{p=0}^{n-1}  \left ( \hat W(n-p)^{ \rho_{n-p}^{(n,2m-d-s)} } e^{-i \hat H_0 \Delta t } \right) \right \vert \vec \Psi(0) \right \rangle. \label{prostatS6.4}
\end{eqnarray}
\setlength{\unitlength}{1.0cm} \begin{figure}  [ht!]
\centering
\resizebox{0.6 \textwidth}{!}{\includegraphics[width=10.5cm,height=10.5cm]{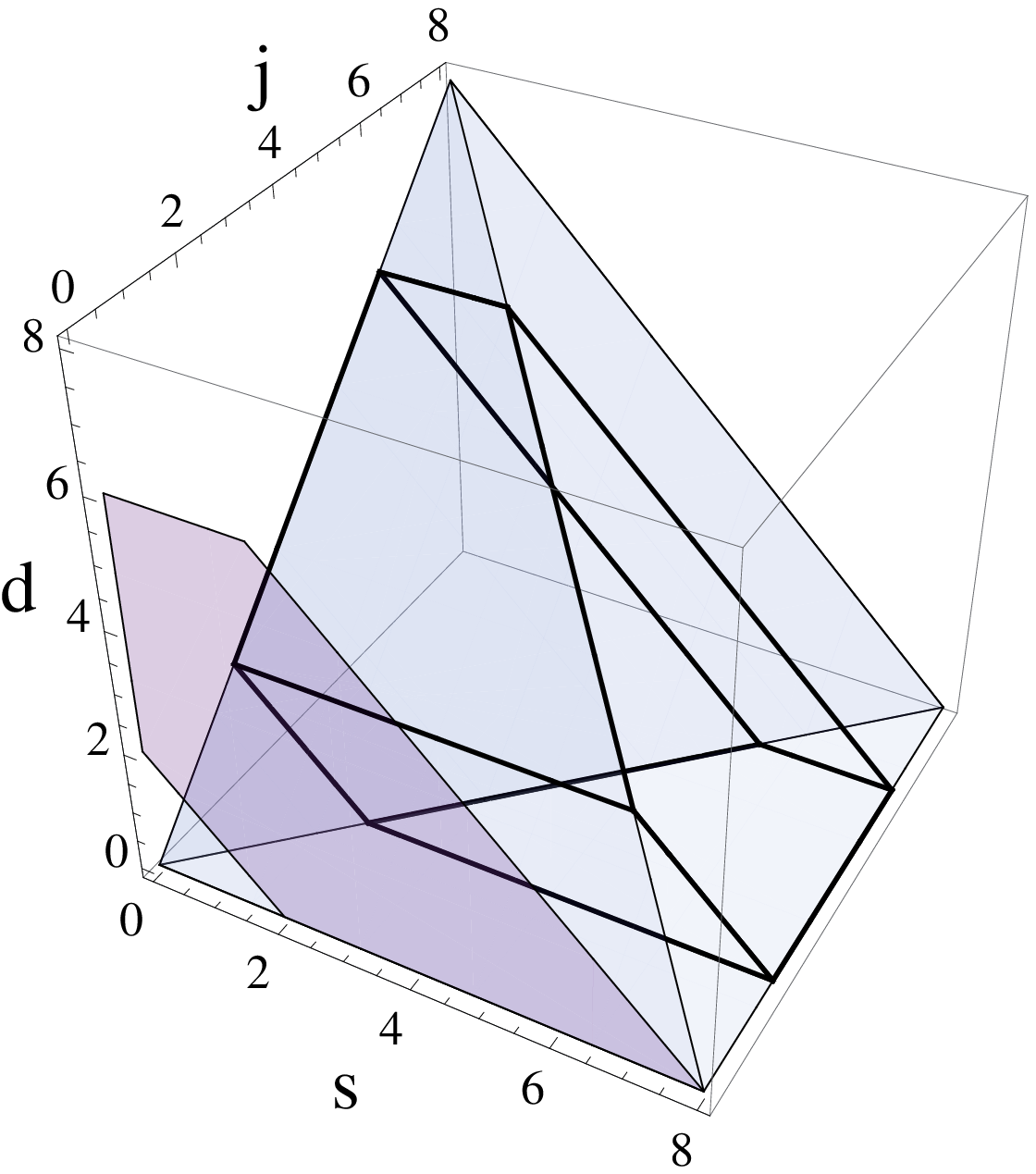}}
\caption{
The frustum of the pyramid presented in figure \ref{fig1}, that we use for the visualization of the threefold sums $\sum_{j = 2m-k}^{k} \sum_{d=0}^j \sum_{s=j-d}^{2m-d}$ on the left and accordingly $\sum_{s = 0}^{2m} \sum_{d=max (2m-k-s, \; 0)}^{min(2m-s, \; k)} \sum_{j= max(2m-k, \; d)}^{min(d+s,\; k)} $ on the right side of (\ref{prostatS6.3}), is shown in this figure for $2m = 8$ and $k=6$ (the edges of this frustum are the highlighted ones). 
Moreover we have drawn in this figure the projection of this body on the $(s,d)$-plane. }
\label{fig4}
\end{figure}
The Eqn. (\ref{prostatS6.3}) can be visualized by the picture that all 3-tuples $(j,d,s)$, which are inside or on the boundary of a frustum of the pyramid discussed above that is the result of a truncation of this pyramid by the two planes $j=2m-k$ and $j=k$, represent exactly all the combinations for $j, d$ and $s$ that are contained both in the threefold sum $\sum_{j = 2m-k}^{k} \sum_{d=0}^j \sum_{s=j-d}^{2m-d}$ on the left side of  (\ref{prostatS6.3}) and in the threefold sum $ \sum_{s = 0}^{2m}  \sum_{d=max(2m-k-s, \; 0)}^{min(2m-s, \; k)} \sum_{j = max(2m-k, \; d)}^{ min(d+s, \; k)}$ on the right side of (\ref{prostatS6.3}), thus these sums are equal. The frustum of the pyramid is presented in figure \ref{fig4} for $2m=8$, $k=6$. The visualization can explained in detail as follows: \newline 
The visualization of the threefold sum on the left side of (\ref{prostatS6.3}) is trivial, because it is easy to see by comparison of the left sides of (\ref{prostatS3.1}) and  (\ref{prostatS6.3}) that the truncation of the pyramid by the planes $j=2m-k$ and $j=k$ only leads to a change of the summation limits of the outer $j$-sum. 
The visualization of the right side of (\ref{prostatS6.3}) is more complicated: \newline
Therefore it is helpful to think first about the projection of the frustum of the pyramid on the $s-d-$plane; 
this projection is presented in figure \ref{fig5} and it implicates that all 2-tuples $(s,d)$ that are inside or on the boundary of this projection represent exactly all the combinations for $s$ and $d$ that are contained in the twofold sum $\sum_{s = 0}^{2m} \sum_{d=max(2m-k-s, \; 0)}^{min(2m-s,k)}$. When we combine this result on the one hand with the question which value has the lower summation limit $j_{min}$ for the innermost sum over $j$ on the right side of  (\ref{prostatS6.3}) for a given 2-tuple $(s,d)$ inside or on the boundary of the projection shown in figure \ref{fig5},  
\setlength{\unitlength}{1.0cm} \begin{figure}  [t]
\centering  
\resizebox{0.6 \textwidth}{!}{\includegraphics[width=10.5cm,height=10.5cm]{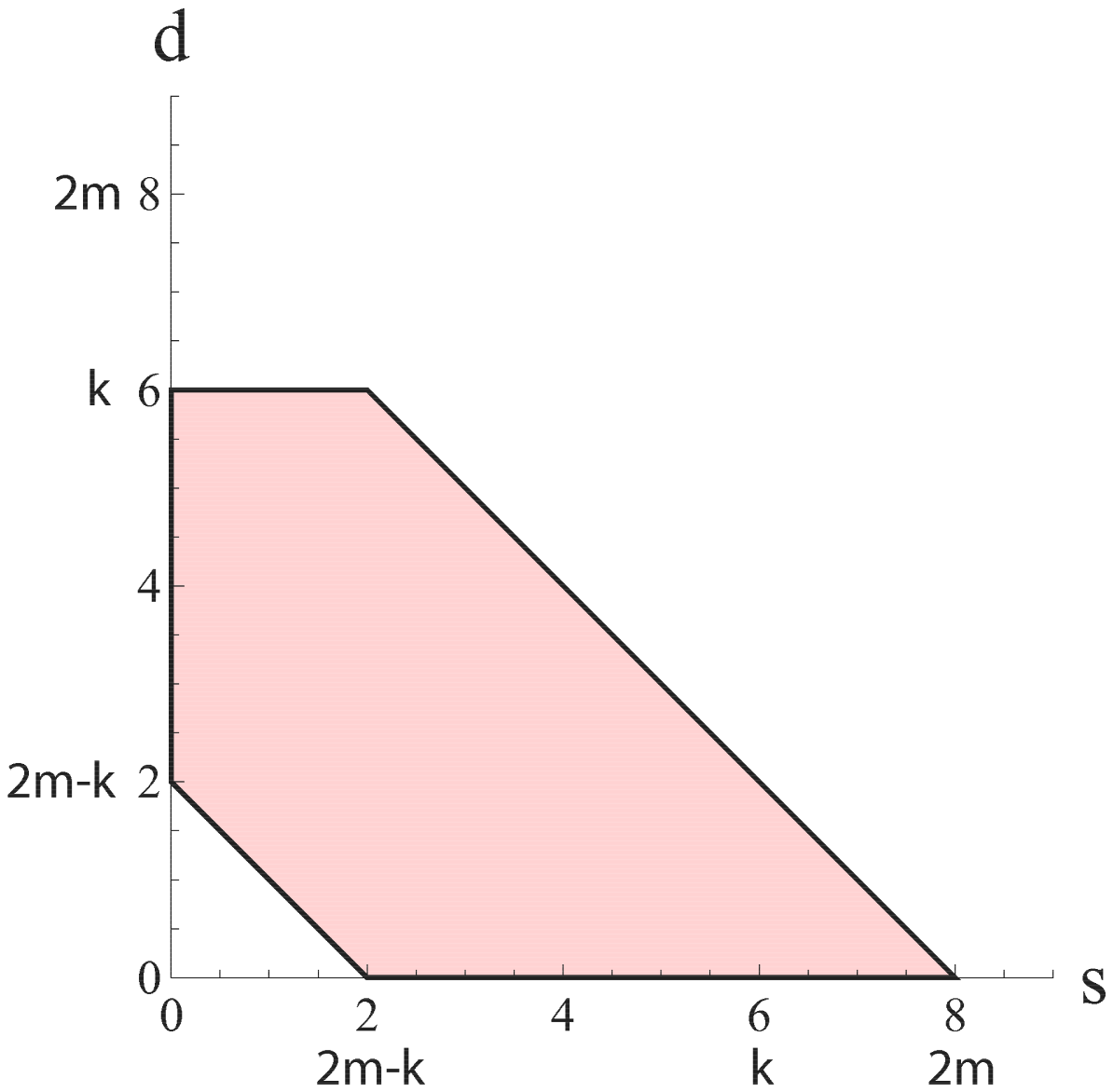}}
\caption{
The polygon, that we use for the visualization of the outer twofold sum $\sum_{s = 0}^{2m}  \sum_{d=max (2m-k-s, \; 0)}^{min(2m-s, \; k)}$  over $s$ and $d$ on the right side of (\ref{prostatS6.3}) is presented here for $2m = 8$, $k=6$. We obtain this polygon by a projection of the frustum of the pyramid shown in figure \ref{fig4} on the $(s,d)$-plane.
}
\label{fig5}
\end{figure}
\setlength{\unitlength}{1.0cm} \begin{figure}  [ht]
\centering  
\resizebox{0.6 \textwidth}{!}{\includegraphics[width=10.5cm,height=10.5cm]{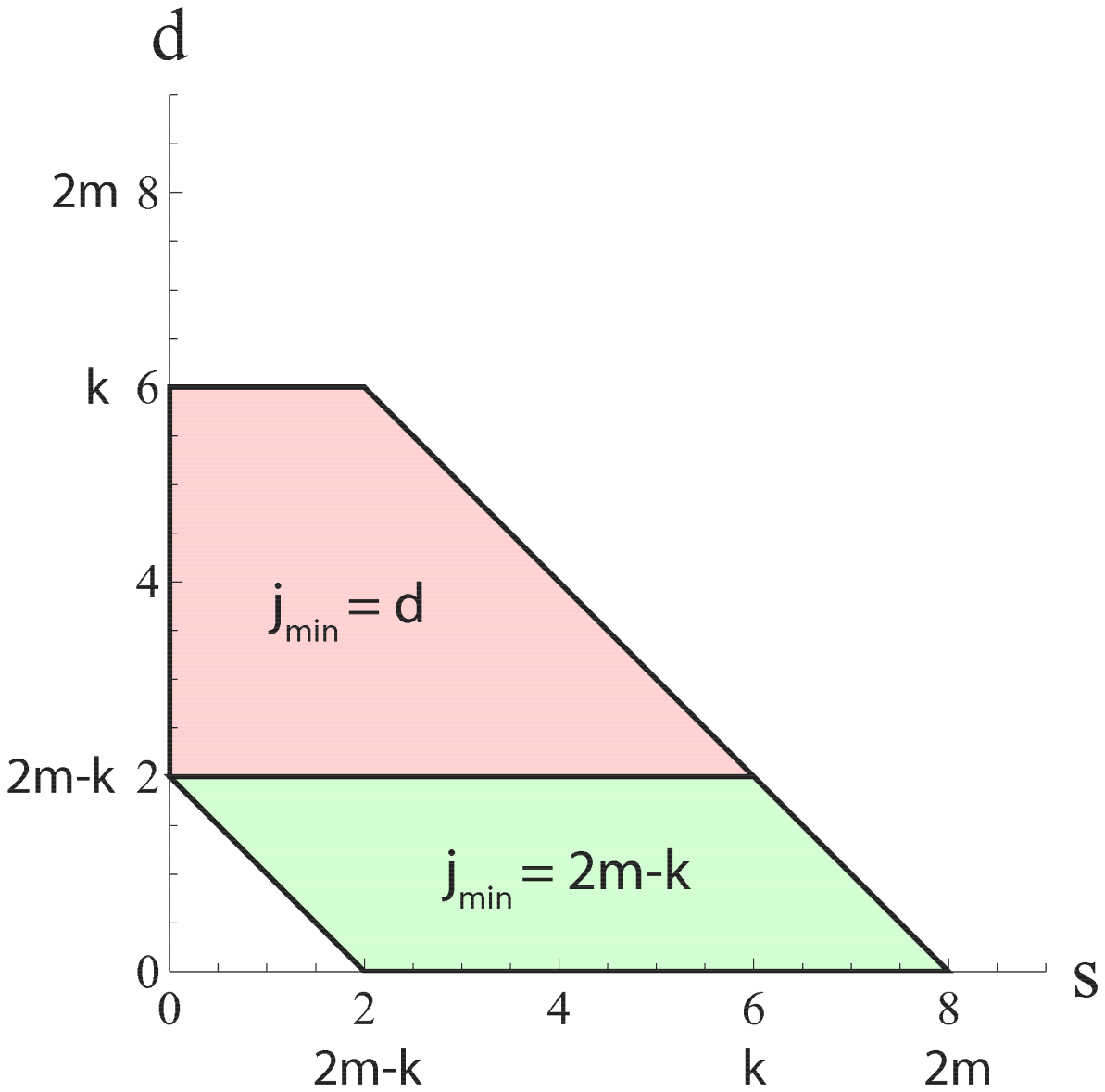}}
\caption{The value for the lower sum limit $j_{min}$ for the innermost sum over $j$ on the right side of (\ref{prostatS6.3})  depends on the values for the summation indices $s$ and $d$ of the two outer sums of this term and moreover on the parameters $2m$ and $k$: For $d < 2m-k$ holds $j_{min} = 2m-k$ and for $ d \geq 2m-k $ holds $j_{min} = d$. The two different areas in the the $(s,d)$-plane for the different kinds of dependencies of $j_{min}$ on $s$ and $d$ can be derived by the form of the frustum of the pyramid shown in figure \ref{fig4}; they are illustrated in this figure for $2m=8$ and $k=6$.} 
\label{fig6}
\end{figure}
\setlength{\unitlength}{1.0cm} \begin{figure}  [ht]
\centering  
\resizebox{0.6 \textwidth}{!}{\includegraphics[width=10.5cm,height=10.5cm]{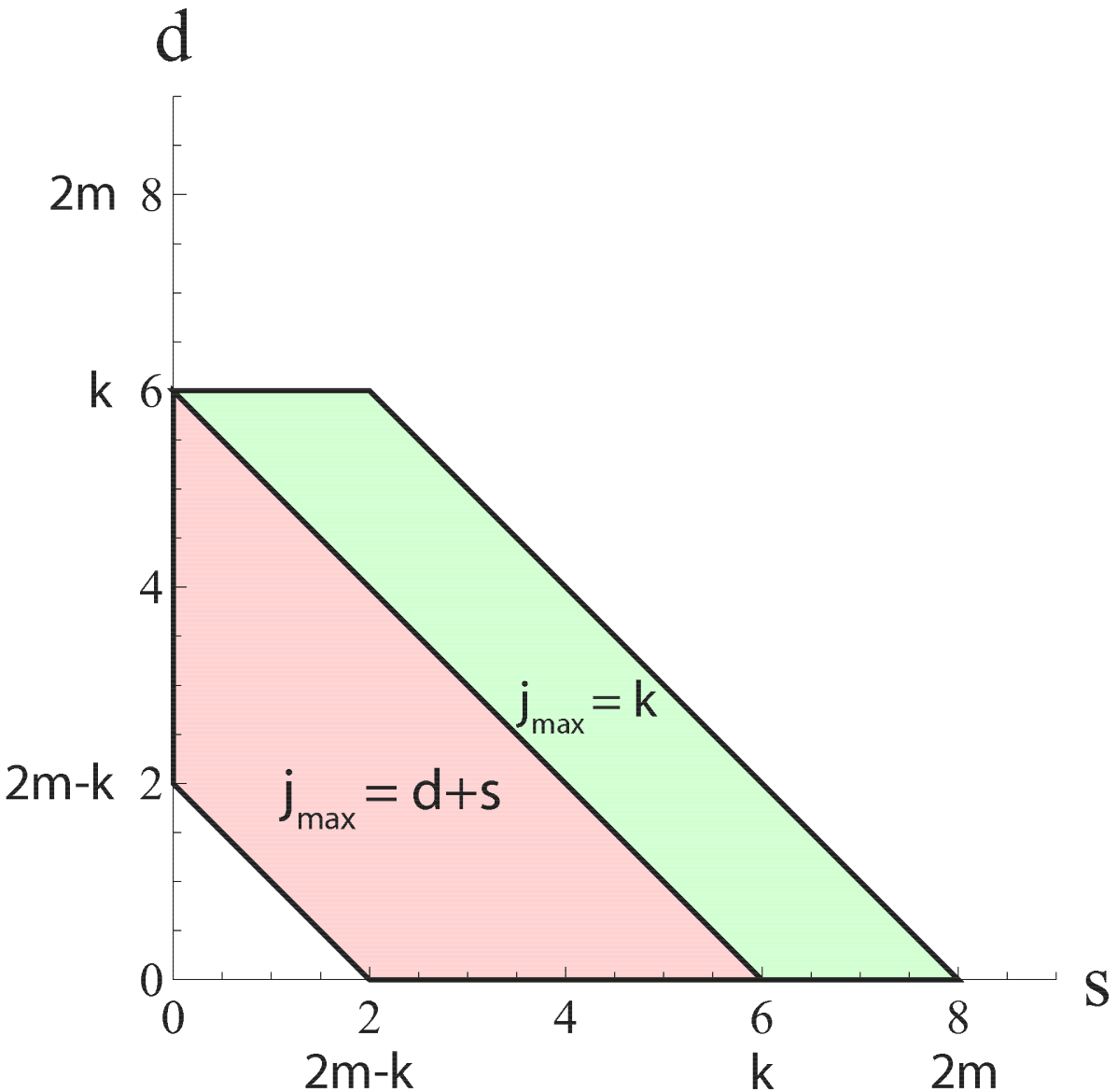}}
\caption{
The value for the higher sum limit $j_{max}$ for the innermost sum over $j$ on the right side of (\ref{prostatS6.3})  depends on the values for the summation indices $s$ and $d$ of the two outer sums of this term and moreover on the parameters $2m$ and $k$: For $d + s \leq  k$ holds $j_{max} = d+s$ and for $ d+s > k $ holds $j_{max} = k$. The two different areas in the $(s,d)$-plane for the different kinds of dependencies of $j_{max}$ on $s$ and $d$ can be derived by the form of the frustum of the pyramid in figure \ref{fig4}; they are illustrated in this figure for $2m=8$ and $k=6$.}
\label{fig7}
\end{figure}
\setlength{\unitlength}{1.0cm} \begin{figure}  [ht]
\centering  
\resizebox{0.6 \textwidth}{!}{\includegraphics[width=10.5cm,height=10.5cm]{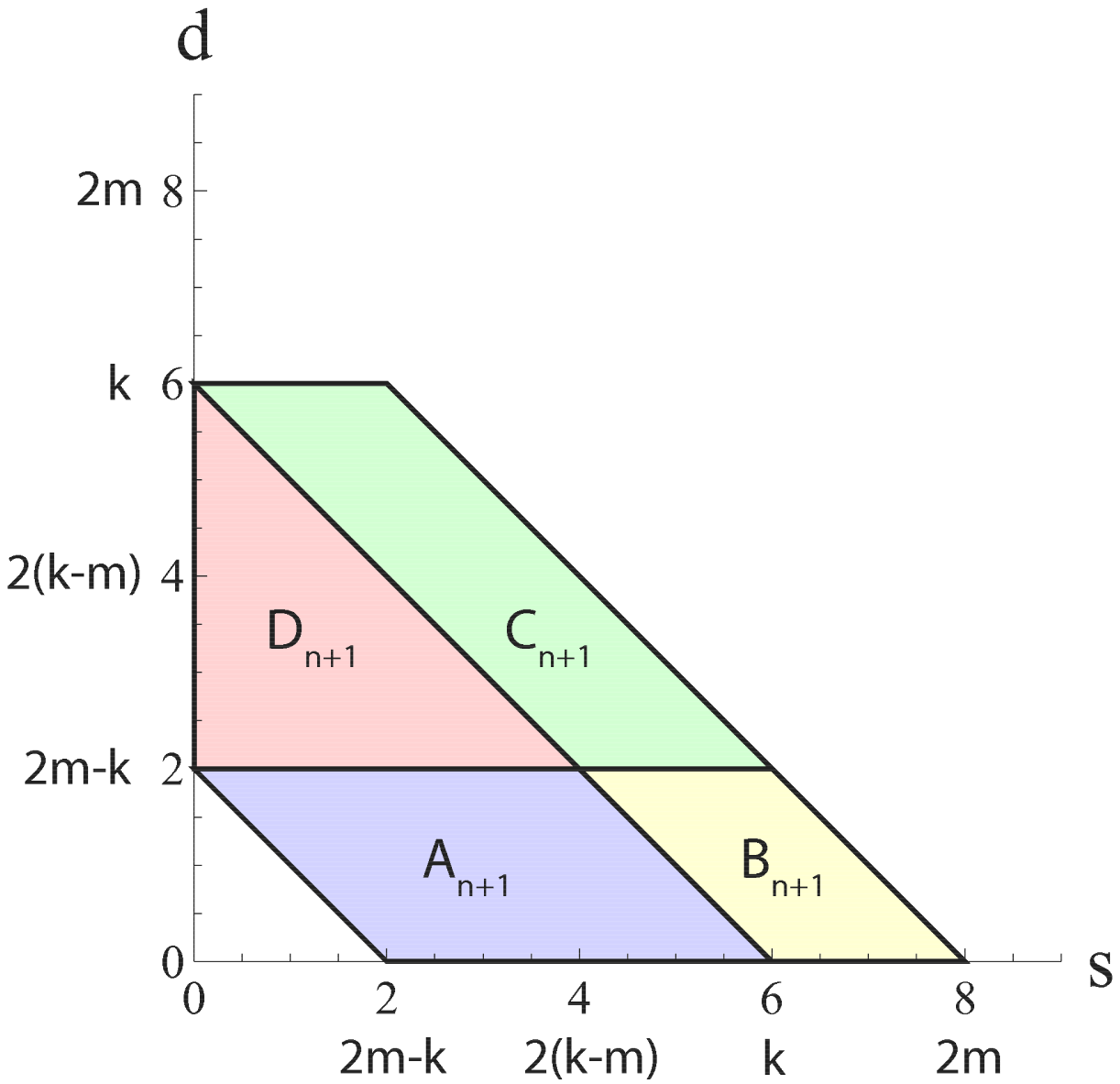}}
\caption{
The corresponding areas inside the polygon presented in figure \ref{fig5} for the four different cases $A_{n+1}$, $B_{n+1}$, $C_{n+1}$ and $D_{n+1}$ presented in the text are shown in this figure for $2m=8$ and $k=6$. (The points being located on boundaries bordering on different areas can arbitrarily allocated to one of these areas allocated cases; the assignment chosen for the definitions of the four cases in the text was made because it makes the calculations easier.) 
}
\label{fig8}
\end{figure}
we realize because of the shape of the frustum of the pyramid shown in figure \ref{fig4}  that $d \leq 2m-k$ implicates $j_{min} = 2m-k$ and  $d > 2m-k$  implicates $ j_{min} = d$ (this result is visualized in figure \ref{fig6}).
On the other hand, when we combine this result with the question which value has the higher summation limit $j_{max}$ for a given 2-tuple $(s,d)$ inside or on the boundary of the projection shown in figure \ref{fig5}, we realize again because of the shape of the frustum of the pyramid shown in figure \ref{fig4} that $d + s \leq k$ implicates $j_{max} = d + s$ and  $d + s > k$  implicates $ j_{max} = k$ (this result is visualized in figure \ref{fig7}).
Thus the innermost sum over $j$ on the left side of  (\ref{prostatS6.3}) is of the form $ \sum_{j = max(2m-k, \; d)}^{ min(d+s, \; k)} $.

Now for (\ref{prostatS6.2}) the summands $\zeta(j,d,s)$ in the threefold sum are again of the special form (\ref{prostatS3.2})
and with (\ref{prostatS3.3}) and (\ref{prostatS3.4}) we get in a fourfold case differentiation the following results for the calculation of the innermost sum over $j$  on the right side of  (\ref{prostatS6.3}), where we name the four cases for afterwards comprehensible systematical reasons $A_{n+1}$, $B_{n+1}$, $C_{n+1}$ and $D_{n+1}$: \newline 
\underline{ $A_{n+1}$: $ d < 2m -k $ and $ d+s \leq k$:} 
\begin{eqnarray}
\hspace{-2,5cm}  \sum_{j = max(2m-k, \; d)}^{ min(d+s, \; k)} (-1)^j \phi (d,s) &=&
\sum_{j = 2m-k}^{d+s} (-1)^j \phi (d,s) = 
(-1)^k \delta_{(d+s+k) \; \textnormal{mod} \; 2, \; 0} \phi (d,s)
\end{eqnarray}
\newline 
\underline{ $B_{n+1}$: $ d < 2m -k $ and $ d+s > k$:} 
\begin{eqnarray}
\hspace{-2,5cm} \sum_{j = max(2m-k, \; d)}^{ min(d+s, \; k)} (-1)^j \phi (d,s) =
\sum_{j = 2m-k}^{k} (-1)^j \phi (d,s) = (-1)^k \phi (d,s)
\end{eqnarray}
\underline{ $C_{n+1}$: $ d \geq 2m -k $ and $ d+s > k$:} 
\begin{eqnarray}
\hspace{-2,5cm} \sum_{j = max(2m-k, \; d)}^{ min(d+s, \; k)} (-1)^j \phi (d,s) =
\sum_{j = d}^{k} (-1)^j \phi (d,s) = (-1)^k \delta_{(k-d) \; \textnormal{mod} \; 2, \; 0} \phi (d,s)
\end{eqnarray}
\underline{ $D_{n+1}$: $ d \geq 2m -k $ and $ d+s \leq  k$:} 
\begin{eqnarray}
\hspace{-2,5cm} \sum_{j = max(2m-k, \; d)}^{ min(d+s, \; k)} (-1)^j \phi (d,s) =
\sum_{j = d}^{d+s} (-1)^j \phi (d,s) = (-1)^d \delta_{ s \; \textnormal{mod} \; 2, \; 0} \phi (d,s) \label{prostatS6.5}
\end{eqnarray}
In figure \ref{fig8} we marked the different areas in the $s-d$-plane for the four different cases $A_{n+1}$, $B_{n+1}$, $C_{n+1}$ and $D_{n+1}$ and the parameter values $2m=8$ and $k=6$. 
Because in  (\ref{prostatS6.4}) appears an sign factor $(-1)^m$, which is not included in the fourfold case differentiation, and in the three cases $A_{n+1}$, $B_{n+1}$, $C_{n+1}$ we get an additional sign factor $(-1)^k$, we have shown for the three cases $A_{n+1}$, $B_{n+1}$ and $C_{n+1}$ that the corresponding terms contributing in the sums on the right side of (\ref{prostatS6.4}) are proportional to a global sign factor $(-1)^{k-m}$. 

Thus we have proven that these terms fulfill the annihilation thesis, so for us only remains the task to show that the terms contributing in the sums on the right side of (\ref{prostatS6.4}) and corresponding to the case $D_{n+1}$, $ d \geq 2m -k $ and $ d+s \leq  k$, fulfill the annihilation thesis, too. 

For the solution of this problem it is practicable to call the sum over all terms on the right side of   (\ref{prostatS6.4}), which fulfill the condition for the case $D_{n+1}$, in our following calculations $D_{n+1,2m, S}^{{(t_{n+1},k)}}$. In this notation, the superscript index $t_{n+1}$ in the brackets denotes that the propagation time $t$ is equal to $(n+1) \; \Delta t$, while the subscript $n+1$ is related to the fact that the case differentiation above, in which the case $D_{n+1}$ appears, is required in the context of a calculation where we wrote out the operators $\hat W(n+1)$ appearing in the bracket term in (\ref{prostatS6.1}) explicitly (see the transformation form 
(\ref{prostatS6.1}) to (\ref{prostatS6.2})).

We can note these terms $D_{n+1,2m, S}^{{(t_{n+1},k)}}$ with the conditions $ d \geq 2m -k $ and $ d+s \leq  k$ and (\ref{prostatS6.5}) in the following way, where we regard that the inequations $ s \leq k-d \leq k - (2m - k) = 2(k-m)$ hold: 
\begin{eqnarray}
&& \hspace{-2,5cm}  D_{n+1,2m, S}^{{(t_{n+1},k)}} = (-1)^m \Delta t^{2m} \sum_{s = 0}^{min \left [ 2(k-m),2m \right ]}  \sum_{d=max(2m-k, \; 2m-k-s, \; 0)}^{min(2m-s, \; k, \; k-s)} (-1)^d \delta_{ s \; \textnormal{mod} \; 2, \; 0} \; \times \nonumber \\
&& \hspace{-1,5cm} \times \; \sum_{\mathcal P_{\vec \nu^{(n,d)}}} \sum_{\mathcal P_{\vec \rho^{(n,2m-d-s)}}}   \left \langle \vec \Psi(0) \left \vert \prod_{q=1}^{n} \left ( e^{i \hat H_0 \Delta t} \hat W(q)^{\nu_q^{(n,d)}} \right ) e^{i \hat H_0 \Delta t} \;   \times \right. \right. \nonumber \\
&& \hspace{-1,5cm}  \times \; \left. \left.  \hat W(n+1)^{s} \; e^{-i \hat H_0 \Delta t} \prod_{p=0}^{n-1}  \left ( \hat W(n-p)^{ \rho_{n-p}^{(n,2m-d-s)} } e^{-i \hat H_0 \Delta t } \right) \right \vert \vec \Psi(0) \right \rangle \label{prostatS6.6}
\end{eqnarray}
Now we consider that for the oscillatory orders holds $k < 2m \leq 2k$, thus we can simplify the max and min functions in the sum limits in the sum over $s$ and $d$ in (\ref{prostatS6.6}) and taking  (\ref{evenW}) into account we get with $s$ substituted by $2r$ as a result:
\begin{eqnarray}
&& \hspace{-2,5cm} D_{n+1,2m, S}^{{(t_{n+1},k)}} = (-1)^m \Delta t^{2m} \sum_{r = 0}^{k-m} W(n+1)^{2r}
\sum_{d=2m-k}^{k-2r} \sum_{\mathcal P_{\vec \nu^{(n,d)}}} \sum_{\mathcal P_{\vec \rho^{(n,2(m-r)-d)}}}  (-1)^d  \; \times \nonumber \\
&& \hspace{-1,5cm} \times \; \left \langle \vec \Psi(0) \left \vert \prod_{q=1}^{n} \left ( e^{i \hat H_0 \Delta t} \hat W(q)^{\nu_q^{(n,d)}} \right )
\prod_{p=0}^{n-1}  \left ( \hat W(n-p)^{ \rho_{n-p}^{(n,2(m-r)-d)} } e^{-i \hat H_0 \Delta t } \right) \right \vert \vec \Psi(0) \right \rangle \label{prostatS6.7}
\end{eqnarray} 
As a next step we rename first $d$ by $j$ and second we write the operator $\hat W(n)$ explicitly out in the same manner like we did this in (\ref{prostatS6.2}) and (\ref{prostatS3}) for the operator $\hat W(n+1)$:
\begin{eqnarray}
&& \hspace{-2,5cm}  D_{n+1,2m, S}^{{(t_{n+1},k)}} = (-1)^m \Delta t^{2m} \sum_{r = 0}^{k-m} W(n+1)^{2r} \left [
\sum_{j=2m-k}^{k-2r} \sum_{d=0}^{j} \sum_{s=j-d}^{2(m-r)-d} (-1)^j  \; \times \right. \nonumber \\
&&  \hspace{-1,5cm}   \left. \times \; \sum_{\mathcal P_{\vec \nu^{(n-1,d)}}} \sum_{\mathcal P_{\vec \rho^{(n-1,2(m-r)-d-s)}}} \left \langle \vec \Psi(0) \left \vert \prod_{q=1}^{n-1} \left ( e^{i \hat H_0 \Delta t} \hat W(q)^{\nu_q^{(n-1,d)}} \right ) e^{i \hat H_0 \Delta t} \;  \times \right. \right. \right. \nonumber \\
&& \hspace{-1,5cm} \left.  \times \;  \left. \left. \hat W(n)^{s} \;  e^{-i \hat H_0 \Delta t} \prod_{p=0}^{n-2}  \left ( \hat W(n-1-p)^{ \rho_{n-1-p}^{(n-1,2(m-r)-d-s)} } e^{-i \hat H_0 \Delta t } \right) \right \vert \vec \Psi(0) \right \rangle \right ]    \label{prostatS6.8}
\end{eqnarray} 
Then we think about (\ref{prostatS6.8}) and realize that there appears the threefold sum $\sum_{j = 2m-k}^{k-2r} \sum_{d=0}^j \sum_{s=j-d}^{2(m-r)-d}$. For a threefold sum of this form and an arbitrary function $\zeta(j,d,s)$ the equation 
\begin{eqnarray}
\hspace{-2,0cm} \sum_{j = 2m-k}^{k-2r} \sum_{d=0}^j \sum_{s=j-d}^{2(m-r)-d} \zeta(j,d,s) =   \sum_{s = 0}^{2(m-r)}  \sum_{d=max(2m-k-s, \; 0)}^{min \left [ 2(m-r)-s, \; k-2r \right]} \sum_{j = max(2m-k, \; d)}^{ min(d+s, \; k-2r)} \zeta(j,d,s) \label{prostatS6.9}
\end{eqnarray}
holds. We can prove (\ref{prostatS6.9}) by substituting the variables $k$ by $k-2r$ and $2m$ by $2(m-r)$ in (\ref{prostatS6.3}) in the upper sum limits. The reader can see for oneself that one can also visualize both sides of (\ref{prostatS6.9}) in a similar way we did this for  (\ref{prostatS6.3}) before. With  (\ref{prostatS6.9}) we can transform (\ref{prostatS6.8}) into 
\begin{eqnarray}
&& \hspace{-2,5cm} D_{n+1,2m, S}^{{(t_{n+1},k)}} = (-1)^m \Delta t^{2m} \sum_{r = 0}^{k-m} W(n+1)^{2r} \; \times \nonumber  \\
&& \hspace{-1,5cm}  \times \; \left \lbrace
\sum_{s=0}^{2(m-r)} \sum_{d=max(2m-k-s, \; 0)}^{min \left [ 2(m-r)-s, \; k-2r \right]} \sum_{j=max(2m-k,\; d)}^{min(d+s, \; k-2r)} (-1)^j \; \times \right. \nonumber \\
&& \hspace{-1,5cm} \left. \times \; \sum_{\mathcal P_{\vec \nu^{(n-1,d)}}} \sum_{\mathcal P_{\vec \rho^{(n-1,2(m-r)-d-s)}}}  \left \langle \vec \Psi(0) \left \vert \prod_{q=1}^{n-1} \left ( e^{i \hat H_0 \Delta t} \hat W(q)^{\nu_q^{(n-1,d)}} \right ) e^{i \hat H_0 \Delta t}  \; \times \right. \right. \right. \nonumber \\
&&  \hspace{-1,5cm}  \left. \times \; \left. \left. \hat W(n)^{s} \; e^{-i \hat H_0 \Delta t} \prod_{p=0}^{n-2}  \left ( \hat W(n-1-p)^{ \rho_{n-1-p}^{(n-1,2(m-r)-d-s)} } e^{-i \hat H_0 \Delta t } \right) \right \vert \vec \Psi(0) \right \rangle \right \rbrace    \label{prostatS6.10}
\end{eqnarray}
and again we get with (\ref{prostatS3.3}) and (\ref{prostatS3.4}) in a fourfold case differentiation the following results for the calculation of the sum over $j$  on the right side of  (\ref{prostatS6.9}), where we now name the four cases $A_{n}$, $B_{n}$, $C_{n}$ and $D_{n}$: \newline 
\underline{ $A_n$: $ d < 2m -k $ and $ d+s \leq k -2r $:} 
\begin{eqnarray}
\hspace{-2,5cm} \sum_{j = max(2m-k, \; d)}^{ min(d+s, \; k - 2r)} (-1)^j \phi (d,s) &=&
\sum_{j = 2m-k}^{d+s} (-1)^j \phi (d,s) = (-1)^k \delta_{(d+s+k) \; \textnormal{mod} \; 2, \; 0} \phi (d,s)
\end{eqnarray}
\newline 
\underline{ $B_n$: $ d < 2m -k $ and $ d+s > k - 2r$:} 
\begin{eqnarray}
\hspace{-2,5cm} \sum_{j = max(2m-k, \; d)}^{ min(d+s, \; k - 2r )} (-1)^j \phi (d,s) =
\sum_{j = 2m-k}^{k - 2r } (-1)^j \phi (d,s) = (-1)^k \phi (d,s)
\end{eqnarray}
\underline{ $C_n$: $ d \geq 2m -k $ and $ d+s > k-2r$:} 
\begin{eqnarray}
\hspace{-2,5cm} \sum_{j = max(2m-k, \; d)}^{ min(d+s, \; k - 2r)} (-1)^j \phi (d,s) =
\sum_{j = d}^{k-2r} (-1)^j \phi (d,s) = (-1)^k \delta_{(k-d) \; \textnormal{mod} \; 2, \; 0} \phi (d,s)
\end{eqnarray}
\underline{ $D_n$: $ d \geq 2m -k $ and $ d+s \leq  k - 2r$:} 
\begin{eqnarray}
\hspace{-2,5cm} \sum_{j = max(2m-k, \; d)}^{ min(d+s, \; k-2r)} (-1)^j \phi (d,s) =
\sum_{j = d}^{d+s} (-1)^j \phi (d,s) = (-1)^d \delta_{ s \; \textnormal{mod} \; 2, \; 0} \phi (d,s) \label{prostatS6.11}
\end{eqnarray}
In an analogue argumentation as in the discussion of  (\ref{prostatS6.4}), in  (\ref{prostatS6.10}) a sign factor $(-1)^m$ appears, which is not included in the calculations done in the fourfold case differentiation. Since in the three cases $A_{n}$, $B_{n}$, $C_{n}$ we get an additional sign factor $(-1)^k$ like for the three cases $A_{n+1}$, $B_{n+1}$, $C_{n+1}$ in the discussion above,  we have shown for the three cases $A_{n}$, $B_{n}$, $C_{n}$ that the corresponding terms contributing in the sums on the right side of (\ref{prostatS6.10}) are proportional to a global sign factor $(-1)^{k-m}$. 

So we have proven that these terms fulfill the annihilation thesis, and now, the task that remains for us is to show that the terms contributing in the sums on the right side of (\ref{prostatS6.10}) and corresponding to the case $D_{n}$, $ d \geq 2m -k $ and $ d+s \leq  k -2r$, fulfill the annihilation thesis, too. 

It's reasonable to call the sum over all terms on the right side of (\ref{prostatS6.10}) which fulfill the condition for the case $D_{n}$ as $D_{n,2m, S}^{(t_{n+1},k)}$ in our following calculations, because the total propagation time is for the term $D_{n,2m, S}^{{(t_{n+1},k)}}$  still $t = (n+1) \Delta t$, but now this term is related to a case differentiation done for a calculation where we wrote out the operator $ \hat W(n)$ explicitly. We can note these terms $D_{n,2m, S}^{(t_{n+1},k)}$ with the conditions $ d \geq 2m -k $ and $ d+s \leq  k - 2r $ and (\ref{prostatS6.10}) in the following way, where we regard that the inequations $ s \leq k - 2r - d \leq k - 2r - (2m - k) = 2(k-r-m)$ hold, and moreover, we substitute for systematical reasons the sum index $r$ by $r_{n+1}$, because $2r$ is the power in which the operator $\hat W(n+1)$ appears in  (\ref{prostatS6.10}): 
\begin{eqnarray}
&& \hspace{-2,5cm}  D_{n,2m, S}^{(t_{n+1},k)} = (-1)^m \Delta t^{2m} \sum_{r_{n+1} = 0}^{k-m} W(n+1)^{2 r_{n+1}} \hspace{-0.3cm} \sum_{s = 0}^{min \left [2(k-r_{n+1}-m), \; 2(m-r_{n+1}) \right]} \hspace{-0.3cm} \delta_{ s \; \textnormal{mod} \; 2, \; 0} \times \nonumber \\
&&  \hspace{-1,5cm}   \times \; \left [ \; \sum_{d=max(2m-k, \; 2m-k-s, \; 0)}^{min \left[2(m-r_{n+1})-s, \; k - 2r_{n+1}, \; k-2r_{n+1}-s \right ]} \hspace{-0.3cm} (-1)^d  \; \times \right. \nonumber \\
&&  \hspace{-1,5cm}  \left. \times \; \sum_{\mathcal P_{\vec \nu^{(n-1,d)}}} \left \langle \vec \Psi(0) \left \vert \prod_{q=1}^{n-1} \left ( e^{i \hat H_0 \Delta t} \hat W(q)^{\nu_q^{(n-1,d)}} \right ) e^{i \hat H_0 \Delta t}  \;  \hat W(n)^{s} \;  e^{-i \hat H_0 \Delta t} \; \times \right. \right. \right. \nonumber \\
&&  \hspace{-1,5cm}  \left. \sum_{\mathcal P_{\vec \rho^{(n-1,2(m-r_{n+1})-d-s)}}}  \left. \left. \prod_{p=0}^{n-2}  \left ( \hat W(n-p-1)^{ \rho_{n-p-1}^{(n-1,2(m-r_{n+1})-d-s)} } e^{-i \hat H_0 \Delta t } \right) \right \vert \vec \Psi(0) \right \rangle \right ] \label{prostatS6.12}
\end{eqnarray}
Now we consider like in the calculation of (\ref{prostatS6.7}) that for the oscillatory orders holds $k < 2m \leq 2k$, thus, we can simplify the max and min functions in the sum limits over $s$ and $d$ in (\ref{prostatS6.12}), and by taking  (\ref{evenW}) into account and renaming $2s$ by $r_n$ and $d$ by $j$, we get as a result for $D_{n,2m, S}^{{(t_{n+1},k)}}$:
\begin{eqnarray}
&& \hspace{-2,5cm}  D_{n,2m, S}^{(t_{n+1},k)} = (-1)^m \Delta t^{2m} \sum_{r_{n+1} = 0}^{k-m} W(n+1)^{2r_{n+1}} \sum_{r_{n} = 0}^{k-r_{n+1}-m} W(n)^{2r_{n}} \; \times \nonumber \\
&& \hspace{-1,5cm} \times \; \left [ \sum_{j=2m-k}^{k-2 \left( r_{n+1}+r_n \right)} (-1)^j  \sum_{\mathcal P_{\vec \nu^{(n-1,j)}}} \left \langle \vec \Psi(0) \left \vert \prod_{q=1}^{n-1} \left ( e^{i \hat H_0 \Delta t} \hat W(q)^{\nu_q^{(n-1,j)}} \right ) \; \times \right. \right. \right. \nonumber \\
&& \hspace{-1,5cm} \left. \sum_{\mathcal P_{\vec \rho^{ \left( n-1,2(m-r_{n+1}-r_{n})-j \right)}}} \left. \left. \prod_{p=0}^{n-2}  \left ( \hat W(n-p-1)^{ \rho_{n-p-1}^{(n-1,2(m-r_{n+1}-r_{n})-j)} } e^{-i \hat H_0 \Delta t } \right) \right \vert \vec \Psi(0) \right \rangle \right ] \label{prostatS6.13}
\end{eqnarray} 
Now we compare (\ref{prostatS6.7}) to (\ref{prostatS6.13}) and realize that we can iterate the implication from (\ref{prostatS6.7}) to (\ref{prostatS6.13}) and thus, get with the abbreviation
\begin{eqnarray}
\Sigma_p =  \left \{ 
\begin{array}{cr}  0 , &  (p \geq n+2)
 \\ \sum_{q=p}^{n+1} r_{q}, & \left (p \in \lbrace 2, \ldots, n+1 \rbrace  \right) \end{array}  \right.  \label{prostatS6.14}
\end{eqnarray}
a general expression for $D_{a,2m, S}^{(t_{n+1},k)}$ for $a=2,3, \ldots, n+1$:
\begin{eqnarray}
&& \hspace{-2,5cm} D_{a,2m, S}^{(t_{n+1},k)} = (-1)^m \Delta t^{2m} \hspace{-0.3 cm} \sum_{r_{n+1}=0}^{k-\Sigma_{n+2}- m} \hspace{-0.2 cm} W(n+1)^{2 r_{n+1}}  \hspace{-0.3 cm} \sum_{r_{n}=0}^{k-\Sigma_{n+1}- m} \hspace{-0.2 cm} W(n)^{2 r_{n}} \cdots \hspace{-0.3 cm} \sum_{r_{a}=0}^{k-\Sigma_{a+1} - m} \hspace{-0.2 cm} W(a)^{2 r_a} \times \nonumber \\
&& \hspace{-1,5cm}  \times \; \left [  \sum_{j=2m-k}^{k-2 \; \Sigma_{a} }  (-1)^j  \sum_{\mathcal P_{\vec \nu^{(a-1,j)}}} \sum_{\mathcal P_{\vec \rho^{ \left( a-1,2 \left (m- \; \Sigma_{a} \right )-j \right)}}} \left \langle \vec \Psi(0) \left \vert \prod_{q=1}^{a-1} \left ( e^{i \hat H_0 \Delta t} \hat W(q)^{\nu_q^{(a-1,j)}} \right ) \; \times \right. \right. \right. \nonumber \\
&&  \hspace{-1,5cm} \left. \times \left. \left. \; \prod_{p=0}^{a-2}  \left ( \hat W(a-1-p)^{ \rho_{a-1-p}^{\left (a-1,2 \left (m - \; \Sigma_{a} \right)-j \right)} } e^{-i \hat H_0 \Delta t } \right) \right \vert \vec \Psi(0) \right \rangle \right ]. \label{prostatS6.15}
\end{eqnarray}
The idea to prove the annihilation thesis is now that if we can prove that one term $D_{a,2m, S}^{(t_{n+1},k)}$ for any value for $a = 2,3, \ldots ,n+1$ is proportional to a global sign factor  $(-1)^{k-m}$, we have accomplished the proof of the annihilation thesis. The reason for this is that because of the results attained in the case differentiations before,  which we did in the implication from the oscillatory order $N_{n+1,2m, S}^{k}$, $k < 2m \leq 2k$ to the term $D_{a,2m, S}^{(t_{n+1},k)}$, the term $N_{n+1,2m, S}^{k}$ is for a proportionality of the term $D_{a,2m, S}^{(t_{n+1},k)}$ to the global sign factor  $(-1)^{k-m}$ proportional to the global sign factor  $(-1)^{k-m}$, too. \newline 
For this task we calculate with (\ref{prostatS6.15}) the term $D_{2,2m, S}^{(t_{n+1},k)}$: 
\begin{eqnarray}
&& \hspace{-2,5cm}  D_{2,2m, S}^{(t_{n+1},k)} = (-1)^m \Delta t^{2m} \hspace{-0.3 cm} \sum_{r_{n+1}=0}^{k-\Sigma_{n+2}- m} \hspace{-0.2 cm} W(n+1)^{2 r_{n+1}} \hspace{-0.3 cm} \sum_{r_{n}=0}^{k-\Sigma_{n+1}- m} \hspace{-0.2 cm} W(n)^{2 r_{n}} \cdots \hspace{-0.3 cm} \sum_{r_{2}=0}^{k-\Sigma_{3}- m} \hspace{-0.2 cm} W(2)^{2 r_{2}} \; \times \nonumber \\
&& \hspace{-1,5cm} \; \times \sum_{j=2m-k}^{k-2 \; \Sigma_{2} }  \left [ (-1)^j  \sum_{\mathcal P_{\vec \nu^{(1,j)}}} \sum_{\mathcal P_{\vec \rho^{ \left( 1,2 \left (m - \; \Sigma_{2} \right )-j \right)}}} \left \langle \vec \Psi(0) \left \vert \prod_{q=1}^{1} \left ( e^{i \hat H_0 \Delta t} \hat W(q)^{\nu_q^{(1,j)}} \right )\; \times \right. \right. \right. \nonumber \\
&& \hspace{-1,5cm}  \; \times \left. \left. \left. \prod_{p=0}^{0}  \left ( \hat W(1-p)^{ \rho_{1-p}^{\left (1,2 \left (m- \; \Sigma_{2}\right)-j \right)} } e^{-i \hat H_0 \Delta t } \right) \right \vert \vec \Psi(0) \right \rangle \right ]. \label{prostatS6.16}
\end{eqnarray}
Moreover, since it is evident that the equations 
\begin{eqnarray}
&& \nu_1^{(1,j)} = j, \; \; \; \;  \rho_1^{\left ( 1,2 \left( m - \Sigma_2 \right ) - j \right)} = 2 \left (m -\Sigma_2 \right ) - j   \nonumber \\
\Longrightarrow &&  \nu_1^{(1,j)} + \rho_1^{\left ( 1,2 \left( m - \Sigma_2 \right ) - j \right)} =  2 \left (m -\Sigma_2 \right )
\end{eqnarray}
hold, thus, we can derive for $D_{2,2m, S}^{(t_{n+1},k)}$ that
\begin{eqnarray}
&& \hspace{-2,5cm} D_{2,2m, S}^{(t_{n+1},k)} = (-1)^m \Delta t^{2m} \hspace{-0.3 cm} \sum_{r_{n+1}=0}^{k-\Sigma_{n+2}- m} \hspace{-0.2 cm} W(n+1)^{2 r_{n+1}} \hspace{-0.3 cm} \sum_{r_{n}=0}^{k-\Sigma_{n+1}- m}  \hspace{-0.2 cm} W(n)^{2 r_{n}}  \cdots \hspace{-0.3 cm} \sum_{r_{2}=0}^{k-\Sigma_{3}- m} W(2)^{2 r_{2}} \times \nonumber \\
&& \hspace{-1,5cm} \times \; W(1)^{2 \left(m-\Sigma_2 \right)} \sum_{j=2m-k}^{k- 2 \; \Sigma_2} (-1)^j  \label{prostatS6.17}
\end{eqnarray}
Then with (\ref{prostatS3.3}) we can carry out the sum over $j$ and get as a final result for $D_{2,2m, S}^{(t_{n+1},k)}$:
\begin{eqnarray}
&& \hspace{-2,5cm} D_{2,2m, S}^{(t_{n+1},k)} = (-1)^{k-m} \Delta t^{2m}  \sum_{r_{n+1}=0}^{k-\Sigma_{n+2}- m} W(n+1)^{2 r_{n+1}} \; \times \nonumber \\ && \hspace{-1,5 cm} \times \; \sum_{r_{n}=0}^{k-\Sigma_{n+1}- m} W(n)^{2 r_{n}} \; \cdots \; \sum_{r_{2}=0}^{k-\Sigma_{3}- m} W(2)^{2r_2} \; W(1)^{2 \left(m-\Sigma_2 \right)} \label{prostatS6.18} 
\end{eqnarray}
We perceive from (\ref{prostatS6.18}) that $D_{2,2m, S}^{(t_{n+1},k)}$ is proportional to a global sign factor $(-1)^{k-m}$, thus, as we discussed before, that means that we have proven for the oscillatory orders $N_{n,2m,S}^k, k< 2m \leq 2k$ the annihilation thesis that these orders are proportional to a global sign factor $(-1)^{k-m}$ and that all terms with an sign factor $(-1)^{k-m+1}$ disappear. $\square$ \newline

\end{appendix}

\end{document}